\documentclass[a4paper,11pt]{article}
\pdfoutput=1 %

\usepackage{jheppub} 
\usepackage{amsmath,amssymb,graphicx,float,slashed,xcolor,multicol}
\usepackage{tabularx}
\usepackage{url}
\usepackage{footmisc}
\usepackage{amsfonts}
\usepackage{cancel}
\usepackage{color}
\usepackage{multirow} 
\usepackage{pifont}
\usepackage{epstopdf}
\usepackage{comment}
\usepackage{booktabs} 
\usepackage{natbib}
\usepackage{array}
\usepackage{mathrsfs}
\usepackage[toc,page]{appendix}
\usepackage{mathtools}
\usepackage{romannum}
\usepackage{bbold}
\usepackage{enumitem}
\usepackage{multirow}
\usepackage{cleveref}
\usepackage{caption,subcaption}
\usepackage{float}
\usepackage{multirow}
\usepackage{upgreek}
\usepackage[toc,page]{appendix}
\usepackage{romannum}
\allowdisplaybreaks
\usepackage{bbm}                
\usepackage{xspace}				
\usepackage{tikz}
\usetikzlibrary{arrows,shapes}
\usetikzlibrary{trees}
\usetikzlibrary{matrix} 
\usetikzlibrary{positioning}				
\usetikzlibrary{calc,through}				
\usetikzlibrary{decorations.pathreplacing}  
\usepackage{pgffor}							
\usetikzlibrary{decorations.pathmorphing}	
\usetikzlibrary{decorations.markings}

\tikzstyle{block} = [draw, rectangle, 
minimum height=3em, minimum width=6em]

\newcommand*{\rom}[1]{\expandafter\@slowromancap\romannumeral #1@}

\def\beq{\begin{equation}}
\def\eeq{\end{equation}}
\def\bea{\begin{eqnarray}}
\def\eea{\end{eqnarray}}

\title{Heavy Long-lived Dark Vector Via a Gluonic Portal}

\author[a]{Xiaoyong Chu,}
\author[a]{Qiyuan Gao,}
\author[b]{Hongkai Liu,}
\author[a]{Teng Ma,} 
\author[a]{Chengjie Yang} 

\affiliation[a]{International Centre for Theoretical Physics Asia-Pacific (ICTP-AP), University of Chinese Academy
of Sciences (UCAS), 100190 Beijing, China} 
\affiliation[b]{High Energy Theory Group, Physics Department,
Brookhaven National Laboratory, Upton, NY 11973, USA}

\emailAdd{
chuxiaoyong@ucas.ac.cn, 
gaoqiyuan23@mails.ucas.ac.cn,
hliu6@bnl.gov,
mateng@ucas.ac.cn, yangchengjie@ucas.ac.cn}

\abstract{We study a dark gauge boson \( Z' \) that exclusively couples to the QCD gluons through higher dimensional operators. These operators are generated from integrating out of heavy ultraviolet   resonances carrying both QCD and dark gauge charges. With \( SU(3)_C \) gauge invariance, charge and parity symmetries preserved, we find  that the leading effective operators are restricted to have the form of \( Z'GGG \) and \( Z'Z'GG \) at dimension-eight, which can naturally render the $Z^\prime$ particle long-lived, and serve as a viable dark matter  candidate. We investigate the phenomenology of these operators with both  collider experiments and cosmological observation, without and with the assumption that this dark gauge boson plays the role of the dominant dark matter component. For an unstable $Z'$, we show that depending on its lifetime, it can be probed by various observables up to ultraviolet physics scale around $10^9$\,GeV. For $Z'$ being dark matter,  we find that $m_{Z^\prime} \gtrsim 1 $~TeV is consistent with the thermal freeze-out scenario. In contrast, in the freeze-in scenario, the extremely small couplings leave the relevant parameter space  largely unconstrained by current experiments.}

\begin{document}
\titlepage	
\maketitle

\flushbottom


\section{Introduction}

Dark sector models typically introduce portals that couple the Standard Model (SM) particles to light dark particles, allowing for non-gravitational probes of the latter. In broken gauge $U(1)_D$ models, the new gauge boson is generally assumed to mix dominantly with the SM hypercharge $U(1)_Y$; see e.g. \cite{Fabbrichesi:2020wbt,  Caputo:2021eaa, Miller:2021ycl} for recent reviews. This portal interaction is minimal, in the sense that it is only given by a renormalizable operator involving only two $U(1)$ field strengths, and characterized by a dimensionless mixing parameter. Due to its simplicity, it has been extensively studied and strongly constrained by a large number of experiments; for recent reviews, see~~\cite{Essig:2013lka, Gori:2022vri, Antel:2023hkf, Cline:2024qzv}. In contrast, portal interactions involving unbroken non-Abelian gauge groups have received much less attention. One early work is Ref.~\cite{Juknevich:2009ji}, which studies the mixing of the SM $Z/\gamma$ with dark gluons. Similarly, Ref.~\cite{Chiu:2022bni} considers mixing between $U(1)_Y$ and dark tensor operators. The goal of our work is to investigate the phenomenology of the reverse setup, where a dark gauge boson couples to the SM gluons, using the effective field theory (EFT) approach. 

The benchmark model features a dark vector field, $Z'$, from either an Abelian $U(1)_D$ or non-Abelian $SU(N)_D$ gauge group, that only interacts with the SM QCD gluons at leading order, e.g. via much heavier intermediate particles charged under both the SM $SU(3)_C$ and the dark gauge groups. Such heavy colored particles are theoretically motivated, for instance, by KSVZ axion models~\cite{Kim:1979if,Shifman:1979if}. Consequently, integrating out these intermediate particles yields effective operators involving only $Z'$ and gluons. 
In addition, direct mixing of $Z'$  with the SM $Z/\gamma$ can be forbidden for non-Abelian $SU(N)_D$, because an unbroken global symmetry in the dark Higgs sector can prevent the mixing, similar to the custodial symmetry in the SM.  Even if direct mixing is not strictly forbidden, like in the $U(1)_D$  case,  it still can be extremely suppressed by  symmetries at high-energy scales (e.g. SUSY partners), or delicate properties of the theory (e.g. brane world models), as studied in \cite{Dienes:1996zr}.  
Moreover, the typical ultraviolet (UV)  models of the gluonic portal operators we study here can automatically suppress this direct mixing, which is generated only at  four-loop level, and thus will be neglected in this work. A quantitative study of the relations between the EFT operators and   concrete UV models is left for future work~\cite{our:followup}.

From now on we focus on the portal interactions between the dark gauge boson $Z'$ and SM gluons. Since the SM $SU(3)_C$ is unbroken, gauge invariance requires at least two gluon fields in the EFT operators. To reduce the operators basis, we impose charge conjugate ({\bf C}) and parity ({\bf P}) symmetries, which is the case when the heavy intermediate particles are vector-like fermions or scalars with real couplings. At last, those intermediate particles are assumed to be much heavier than the $Z'$ mass and are not involved in the breaking of dark gauge symmetry. That is, the EFT operators introduced do not contain any dark Higgs boson, and they must be expressed in terms of field strengths of dark gauge boson and gluons at the leading order. For more general cases, the complete set of EFT bases for dark gauge boson can be found in~\cite{Dong:2024dce, Liu:2023jbq, Dong:2022jru}. 
As a result, dimension-4 and dimension-6 operators are forbidden (see e.g., Ref.~\cite{Dong:2024dce}).\footnote{Dimension-6 mixing between $Z'$ and the gluons is allowed for parity-odd interactions~\cite{Bramante:2011qc} and  pseudo-vector $Z'$~\cite{Alwall:2012np}. While not considered here, intermediate particles charged under electroweak gauge groups would allow mixing between $Z'$ and the $Z$ boson,  through CP-violating dimension-4 operators, e.g. ~\cite{Chang:1988fq,Keung:2008ve},  or CP-conserving dimension-6 operators, e.g. ~\cite{Dudas:2013sia, Ducu:2015fda}. }
The lowest dimension of the allowed effective portal operators is eight, in terms of $Z'GGG$ and $Z'Z'GG$. A key implication of these simplified assumptions is that the $Z'$ can naturally be long-lived, as its decay is mediated at most by dimension-8 operators. As a comparison, a dark scalar $S$ typically decays via the dimension-5 operator $SGG$. Therefore, stabilizing $S$ requires an explicit ${\mathcal Z}_2$ symmetry~\cite{Godbole:2015gma, Godbole:2016mzr}, and we assume that it always decays quickly and does not sizably affect those observables considered below.
 
This paper is organized as follows.  The complete set of leading-order effective operators of the form $Z'GGG$ and $Z'Z'GG$ is presented in Sec.~\ref{sec.operator}. In Sec.~\ref{sec.pheno}, we derive the collider and cosmology constraints on the $Z'$ in the general case. Then we investigate the scenario where the $Z'$ boson serves as the dominant dark matter (DM) component and the corresponding constraints from the relic abundance and direct/indirect detections are analyzed in Sec.~\ref{sec.dm}, considering both freeze-in and freeze-out production mechanisms. Sec.~\ref{sec:conclu} is devoted to our final conclusions. 
In addition, the Appendices include the derivation of the decay widths induced by the EFT operators, explicit expressions of relevant cross sections, among other detailed calculations.

\section{Leading Operators between Dark Gauge Bosons and Gluons}\label{sec.operator} 

In this section, we explore the relevant portal interactions within the EFT framework. Our analysis focuses on the EFT that arises from a class of UV completions in which the heavy states are charged under both the dark gauge group and QCD $SU(3)_C$ group. These heavy states thus generate portal interactions between $Z'$ and gluons, without introducing any breaking of the dark gauge symmetry. 
Therefore, the EFT operators at leading order from these states are dark gauge invariant.\footnote{Although dark Higgs may induce symmetry-breaking effects to this portal at more-than-two-loop order, the associated  effects are very subleading, given that the vacuum expectation value of the dark Higgs is much lighter than intermediate state masses. For general cases, those Wilson coefficients of the dark symmetry-violating EFT operators are not independent and can determined by unitarity in a UV completion~\cite{Liu:2023jbq}.}  Therefore, in the following, we only list the EFT operators composed of dark gauge boson field strength $Z^\prime_{\mu \nu} $.  Therefore, the leading-order operators only appear at dimension-8. Furthermore, if {\bf C} and {\bf P} symmetries are preserved at UV scale, there are only six independent ones~\cite{our:followup}. For simplicity, we limit ourselves to these operators. 

First of all, we introduce the operators with an explicit ${\mathcal Z}_2$ symmetry for the $Z'$ particle, $Z^\prime Z^\prime GG$. A full set of independent effective operators at dimension-8 have been listed in \cite{Ellis:2018cos, Dong:2024dce,Liu:2023jbq,Dong:2022jru}, yielding 
\begin{equation}
\begin{aligned}
	{\mathcal O}_{1} &=& \frac{X_1}{\Lambda^4}  \,Z'_{\alpha\beta}  Z'^{\alpha\beta} \times {\rm Tr}[G_{\mu\nu}  G^{\mu\nu}]\, ,\\
	{\mathcal O}_{2} &=&  \frac{X_2}{\Lambda^4}  \,Z'_{\mu\beta}  Z'^{\alpha\nu} \times {\rm Tr}[G_{\alpha\nu}  G^{\mu\beta}]\, ,\\
   {\mathcal O}_{3} &=&  \frac{X_3}{\Lambda^4}  \,Z'_{\nu\beta}  Z'^{\alpha\nu} \times {\rm Tr}[G_{\alpha\mu}  G^{\mu\beta}]\, ,\\
	{\mathcal O}_{4} &=&  \frac{X_4}{\Lambda^4}  \, Z'_{\mu\beta}  Z'_{\nu \alpha} \times {\rm Tr}[G^{\alpha\mu}  G^{\beta\nu}]\, , 
    \label{eq:ZZGG}
    \end{aligned}
\end{equation}
where $G_{\mu\nu}= G_{\mu\nu}^c T^c$ with $G_{\mu\nu}^c$ being the gluon field strength, $T^c$ being  $SU(3)_C$ generators in the fundamental representation ($c = 1,2,..,8$).  Hereafter this kind of gauge vector boson is referred to as a \textit{gluonic $Z^\prime$}.

For the operators with one $Z^\prime$ and three gluons, the dark gauge boson field strength can contract with the Lorentz indices of   three gluon field strengths in only two distinct ways~\cite{Stohr:1993uj,Dong:2024dce}, which are simply
\begin{equation}\label{eq:ZGGG}
\begin{aligned}
	{\mathcal O}_{5} &=&\frac{Y_1}{\Lambda^4}  \,  Z'_{\nu\alpha} \times {\rm Tr}[G^{\alpha\beta} G_{\beta\mu} G^{\mu\nu}]  \, ,\\
	{\mathcal O}_{6} &=& \frac{Y_2}{\Lambda^4} \,  Z'_{\alpha\beta} \times   {\rm Tr}[G^{\alpha\beta} G_{\mu\nu} G^{\mu\nu}  ] \, .
    \end{aligned}
\end{equation}
The EFT amplitudes of such operators have been studied, where only the symmetric color factor would appear at dimension-8 \cite{Shadmi:2018xan}.
In addition, we have checked that  $Z G D^2  G $  would not add any additional independent terms to those operators above. This can be directly seen from the equality $D^\lambda D^\alpha -   D^\alpha D^\lambda = -ig_s G^{\lambda\alpha}$, where $g_s$ is the QCD coupling constant. This is also confirmed by previous works on the on-shell EFT construction~\cite{Dong:2024dce,Liu:2023jbq,Dong:2022jru}.     

The operators $O_5$ and $O_6$ enable the $Z'$ particle to decay into three jets if its mass, $m_{Z'}$, is above the $\mathcal{O}(1)$ GeV scale. Nevertheless, the two operators can be eliminated if a residual ${\mathcal Z}_2$ symmetry survives after dark gauge symmetry breaking. This can be naturally achieved in various ways, for instance those originating from non-Abelian gauge groups, as already demonstrated in many vector DM  models~\cite{Birkedal:2006fz,  Hambye:2008bq, Hambye:2009fg, Diaz-Cruz:2010czr, Bhattacharya:2011tr, Baek:2012se, Farzan:2012hh, Carone:2013wla, Chen:2014cbt, DiChiara:2015bua, Arcadi:2017jqd, Saez:2018off, Ko:2020qlt, Cai:2021wmu, Alonso-Alvarez:2023rjq}. 
Therefore, by omitting  $O_5$ and $O_6$ to make $Z'$ absolutely stable, we also analyze its corresponding DM phenomenology below.  Nevertheless, we find that even in the presence of such operators the gluonic $Z'$ can still serve as the DM candidate in the freeze-in scenario,  whereas the freeze-out scenario apparently requires their absence.

\section{General Phenomenology of the Gluonic $Z^\prime$} \label{sec.pheno}
As established in the preceding analysis, the gluonic \( Z' \) couples to SM particles exclusively through the dimension-8 EFT  operators given in Eqs.~(\ref{eq:ZZGG}-\ref{eq:ZGGG}), featuring the characteristic interaction terms \( Z'Z'GG \) and \( Z'GGG \), respectively. This section is devoted to investigating its   potential signatures, mostly induced by the longevity and the final decay of $Z'$,  at colliders, and in cosmological and astrophysical experiments. To illustrate the results in a clear manner,  in this section we  only derive quantitative bounds for a combination of $\mathcal {O}_1$ and  $\mathcal {O}_5$ operators, where $X_1 = Y_1 =1$ and all other Wilson coefficients are set to vanish,
thus experimental data are used to put constraints on the associated UV scale, $\Lambda$. 
For general UV models, $X_{1-4}$ and $Y_{1,2}$ should all appear, being of the same order  in practice. Nevertheless, given the high exponent on $\Lambda$, the corresponding constraints can at most change by a factor of two when they are re-scaled to apply to general models. 

\subsection{The decay  of the gluonic $Z^\prime$}   

\begin{figure}
    \centering
    \includegraphics[width=0.495\linewidth]{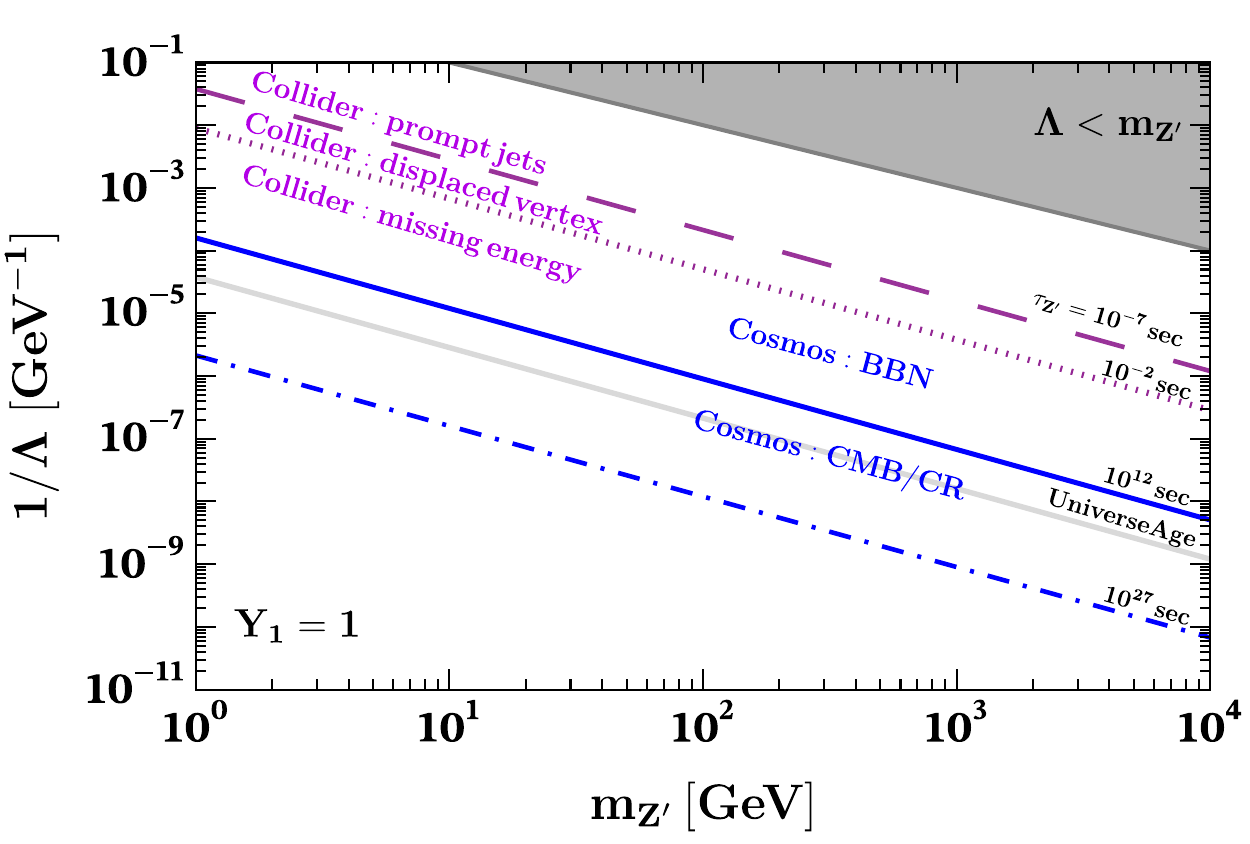}
    \includegraphics[width=0.495\linewidth]{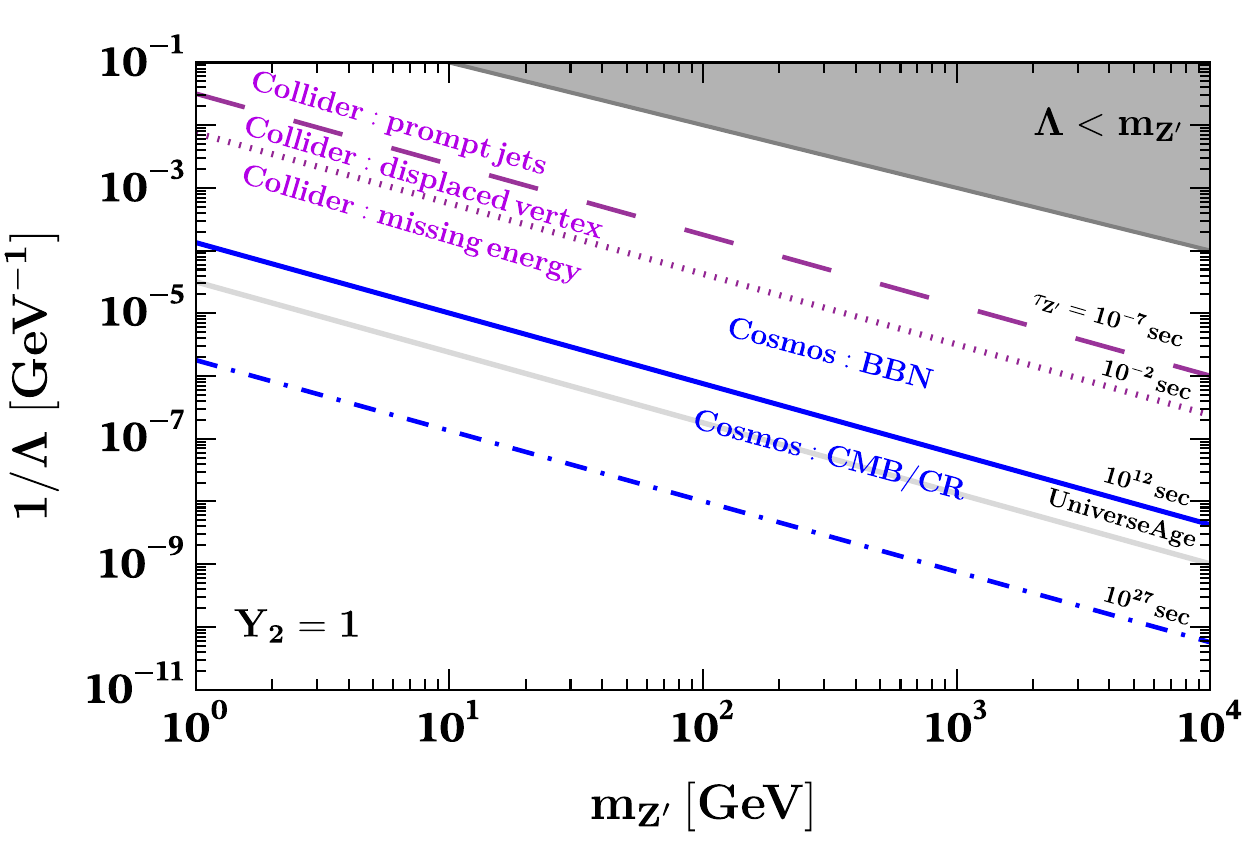}
    \caption{Dependence of the $Z'$ proper lifetime on the model parameters and its impact on experimental signatures, shown for $Y_1 = 1$ ({\bf left} panel) and $Y_2 = 1$ ({\bf right} panel). Diagonal lines indicate iso-lifetime contours (e.g., \(\tau_{Z'} = 10^{-7}\,\mathrm{sec},\, 10^{-2}\,\mathrm{sec},\, 10^{12}\,\mathrm{sec},\, 10^{27}\,\mathrm{sec}\)), representing typical lifetimes that are sensitive to different experiments: purple for collider probes and blue for cosmological probes. These contours partition the plane into several search regimes: prompt decays for \(\tau_{Z'}<10^{-7}\,\mathrm{sec}\); displaced vertices for \(10^{-7}\,\mathrm{sec}\lesssim\tau_{Z'}\lesssim10^{-2}\,\mathrm{sec}\); and missing-energy channels for \(\tau_{Z'}\gtrsim10^{-7}\,\mathrm{sec}\). Cosmological bounds apply to \(10^{-2}\,\mathrm{sec}\lesssim\tau_{Z'}\lesssim10^{12}\,\mathrm{sec}\) (BBN) and \(10^{12}\,\mathrm{sec}\lesssim\tau_{Z'}\lesssim10^{27}\,\mathrm{sec}\) (CMB/CR). The gray solid line indicates the age of the Universe,   serving as a reference for whether the $Z'$ is effectively stable as a DM candidate. The shaded gray region with \(\Lambda<m_{Z'}\) marks the breakdown of the EFT approach. }
    \label{fig:lifetime}
\end{figure}

We start with the decay width of $Z'$, as its value is crucial for determining the experimental signatures. For the parameter region studied in this work, $Z'$ dominantly decays into three gluons, with its decay width determined by the operators $\mathcal{O}_5$ and $\mathcal{O}_6$ in Eq.~\eqref{eq:ZGGG}: 
\begin{equation}\label{eq:decay}
\Gamma_{Z^{\prime}} = \frac{m_{Z^{\prime}}^9}{41472 \pi^3 \Lambda^8}\left(2 Y_1^2 + 7 Y_1 Y_2 + 8 Y_2^2\right).
\end{equation}
In turn, the proper decay length is given by
\begin{equation}\label{zlifetime}
L_{Z'}\equiv c\tau_{Z'}  \simeq  1\,\text{meter} \left(\frac{2}{2Y_1^2 + 7Y_1 Y_2 + 8Y_2^2}\right) \left(\frac{1~\text{TeV}}{m_{Z'}}\right)^9 \left(\frac{\Lambda}{40\,\text{TeV}}\right)^8.
\end{equation}
where the at-rest lifetime $\tau_{Z'}=1/\Gamma_{Z^{\prime} }$. Fig.~\ref{fig:lifetime} illustrates the representative benchmark values relevant for collider phenomenology (in purple) and cosmology (in blue). In the shaded gray region of the figure there exists $\Lambda < m_{Z'}$, where  the EFT approach does not apply.

At the parton level, the three-gluon decay channel is dominant for a $Z^\prime$ boson with mass above 1\,GeV, yielding the  signatures studied below. Other potentially detectable decay channels are subject to additional suppression. For instance,  $Z' \to g g \gamma$ only happens through  even higher-dimensional EFT operators within our setup, and thus is much less likely. Channels like $Z' \to h \gamma$ and $Z' \to \bar{f} f$ are further suppressed, as there is no intermediate beyond the SM (BSM) particles connecting $Z'$ to electroweak interactions at tree level~\cite{Bernreuther:1989rt}. At last, $Z' \to \gamma \gamma$ is forbidden by the Landau-Yang theorem~\cite{Landau:1948kw,Yang:1950rg,Chang:1988fq,Keung:2008ve}.  

\subsection{Collider Constraints}
\label{sec:collider}

As is shown by Eq.~\eqref{zlifetime} and Fig.~\ref{fig:lifetime},  the gluonic $Z'$ boson can have a macroscopic decay length in colliders  for TeV-scale $\Lambda$, leading to detectable signatures, such as missing energy and displace vertex events. This section studies such signatures at the Large Hadron Collider (LHC).

\subsubsection{Missing Energy and Displaced Vertex}\label{sec.me_dv}

\begin{figure}[t]
	\centering
	{\includegraphics[width=0.23\linewidth]{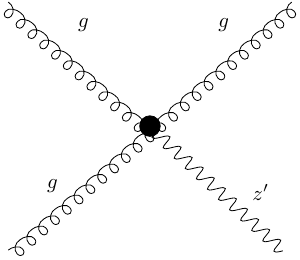}}
    {\includegraphics[width=0.17\linewidth]{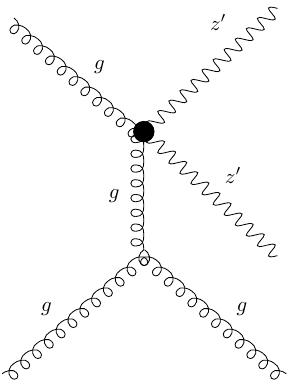}}
    {\includegraphics[width=0.29\linewidth]{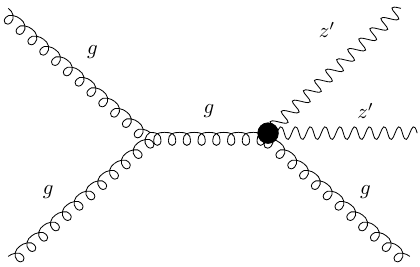}}
    {\includegraphics[width=0.23\linewidth]{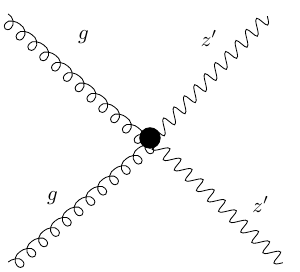}}
	\caption{The representative leading-order Feynman diagrams contributing to the $Z'$ production at LHC, induced by the operators $\mathcal{O}_{1-6}$. The operators $\mathcal{O}_5$ and $\mathcal{O}_6$ generate a $gggZ^\prime$ vertex corresponding to the leftmost diagram. The four operators $\mathcal{O}_1$ to $\mathcal{O}_4$ generate a $ggZ^\prime Z^\prime$ vertex contributing to the three rightmost diagrams.}	
	\label{fig:feyndia}
\end{figure}

To produce  gluonic $Z'$ bosons at LHC,  the dominant channels are the  $pp \to Z'Z'(j)$ and $pp \to Z' j$ processes, induced by the operators $\mathcal{O}_{1-4}$ and $\mathcal{O}_{5,6}$, respectively. The representative Feynman diagrams are given in Fig.~\ref{fig:feyndia} to yield  missing–energy events (the three leftmost ones) and displaced vertex (the first and fourth ones) events. This is due to the fact that the former signal requires at least one visible energetic jet.

Those \( Z' \) bosons produced at LHC with a sufficiently long decay length can lead to an observable event with a displaced vertex or missing transverse energy  \( E_T^{\text{miss}} \). To characterize the detected signal, we introduce  $\mathcal{P}_D$ to describe the probability that the $Z'$ decays within the length interval $[L_1, L_2]$:    
\begin{equation}
    \mathcal{P}_D(L_1,L_2)=
      \left[
        \exp\!\left(-\frac{L_1\, m_{Z'}}{p_{Z'} L_{Z'}}\right)
        - \exp\!\left(-\frac{L_2\, m_{Z'}}{p_{Z'} L_{Z'}}\right)
      \right],
\end{equation}
where $L_{Z'}$ is the proper decay length defined above, $p_{Z'}$ is the momentum of $Z'$ in the laboratory frame, which is typically around $\mathcal{O}(1)$~TeV. For demonstration, we take $(L_1, L_2) = (3\,{\rm m}, 14\,{\rm m})$ for  displaced vertex events, and $(14\,{\rm m}, \infty)$ for the missing transverse energy events. Naively speaking, only particles with lifetimes of $\tau_{Z'} \gtrsim 10^{-7}$\,s have a non-negligible probability to escape the detector, producing a genuine \( E_T^{\text{miss}} \) signature. For shorter lifetimes, the dominant constraints instead arise from displaced-vertex searches.

\vspace{0.12cm}

To derive the bounds from the missing energy search, we adopt a recent ATLAS study that performed a dedicated search for signatures with an energetic jet and missing transverse energy,  using \( 139 \, \text{fb}^{-1} \) data at a center-of-mass energy of \( \sqrt{s} = 13 \, \text{TeV} \)~\cite{ATLAS:2021kxv}.  As no significant deviation from the SM predictions has been observed, the ATLAS collaboration obtains upper limits on the signal cross section for various $E_T^{\text{miss}}$ thresholds. Following the same analysis strategy, we impose the selection criterion on the missing energy \( E_T^{\text{miss}} > 1200~\mathrm{GeV} \), and demand exactly one central jet within the detector acceptance, $|\eta_j|<2.4$, in order to optimize the signal sensitivity for our  dimension-8 operators.  The final signal cross-section can be expressed as 
\begin{equation}
    \sigma_{\text{sig}}
    = \sigma_{\text{prod}}\times \mathcal{A}\times \epsilon\times \mathcal{P}_D(L_1,L_2)^n,
\end{equation}
where $\sigma_{\text{prod}}$ is the production cross section, $\mathcal{A}$ denotes the signal acceptance determined by the selection cuts and the exponent $n$ is the number of $Z'$ bosons in the final state. For the detection efficiency, we simply adopt \(\epsilon = 1\), as our   bounds on \(\Lambda\) are insensitive to modest variations of $\epsilon$ due to the \(\Lambda^8\) dependence in the cross-sections, as already stated above. 
We simulate the signal production processes at leading order using \path{MadGraph5_aMC@NLO}~\cite{Alwall:2011uj}
and \texttt{FeynRules}~\cite{Christensen:2008py}. 

\begin{figure}[t]
    \centering
    \includegraphics[width=0.495\linewidth]{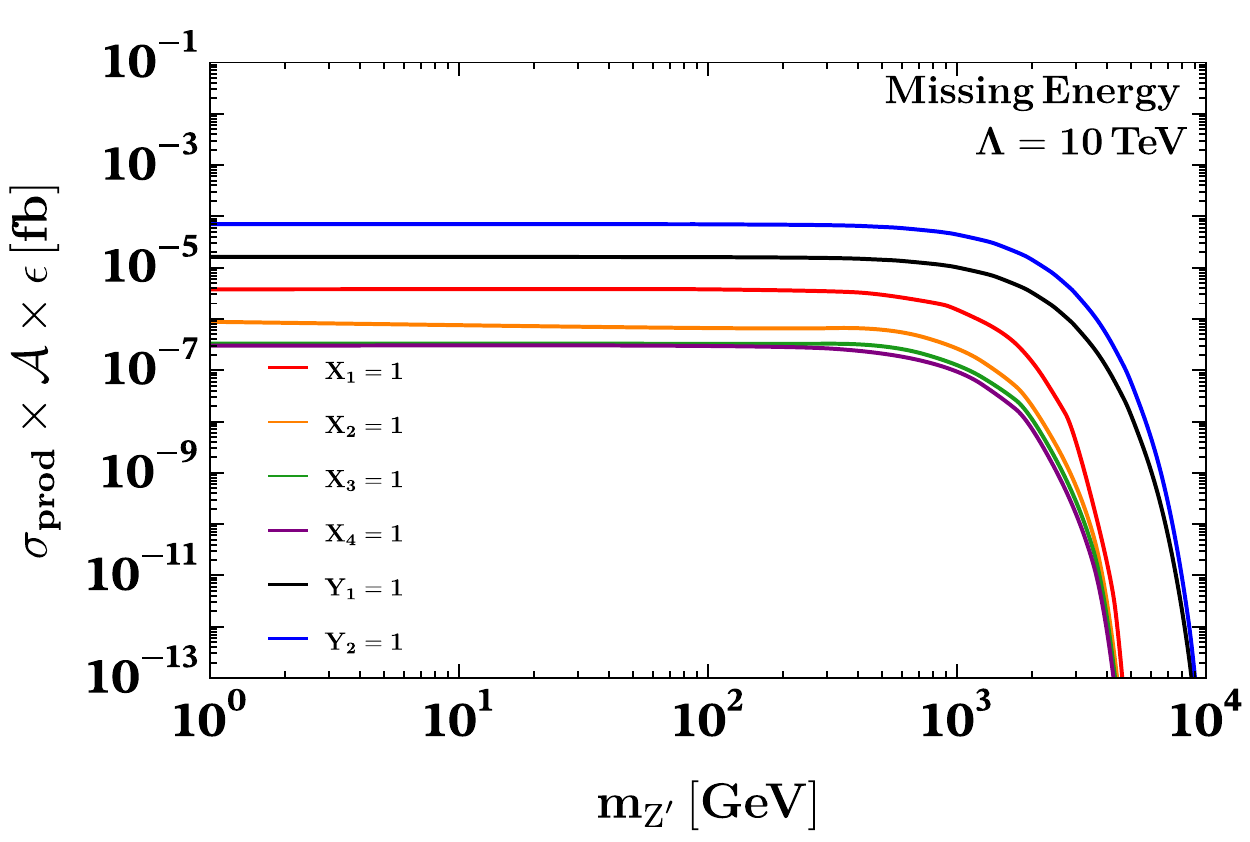}
    \includegraphics[width=0.495\linewidth]{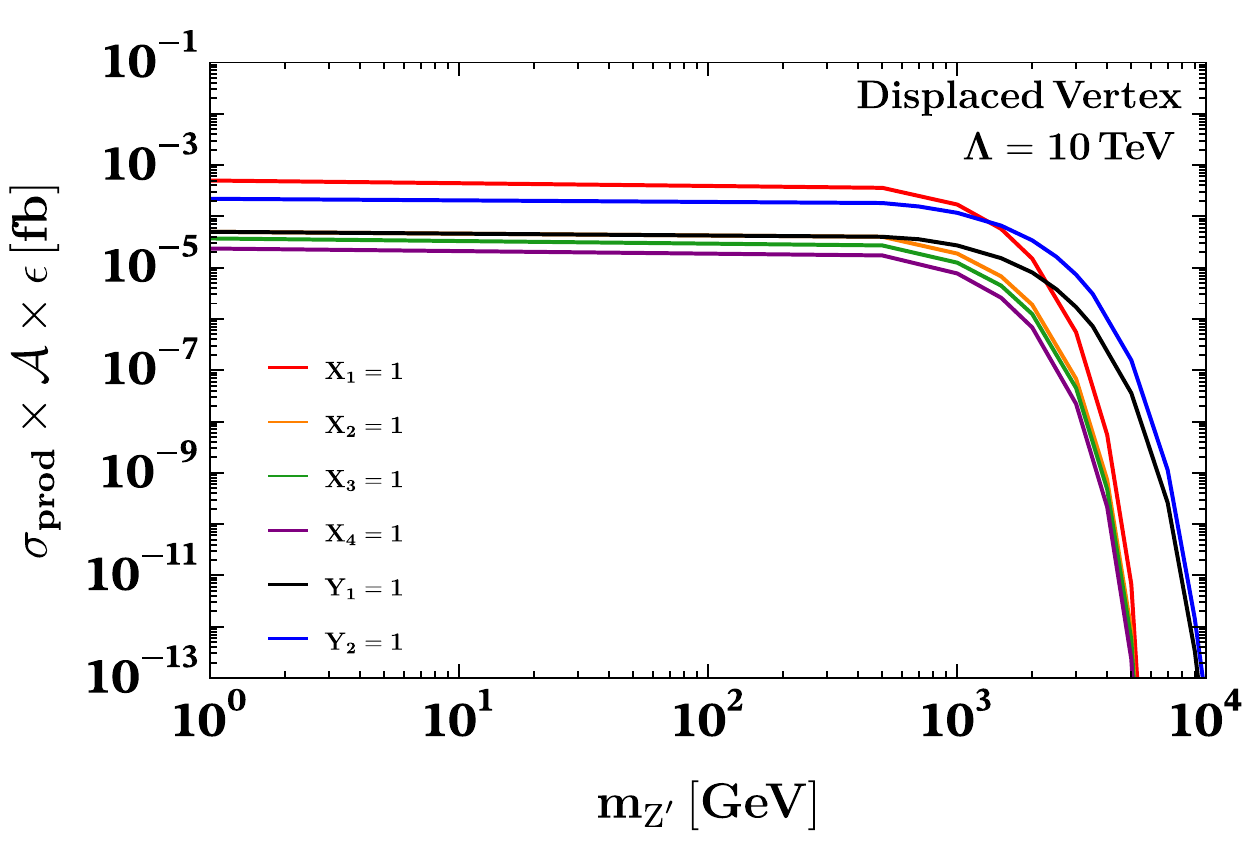}
    \caption{The quantities $\sigma_{\mathrm{prod}} \times \mathcal{A} \times \epsilon$ for the missing energy ({\bf left} panel) and displaced vertex ({\bf right} panel) signatures at the ATLAS detector with $\sqrt{s} = 13~\mathrm{TeV}$, evaluated at a fixed cutoff scale $\Lambda = 10~\mathrm{TeV}$. For the missing energy signature, we impose the selection criteria $E_T^{\text{miss}} > 1200~\mathrm{GeV}$, $|\eta_j| < 2.4$, and $\epsilon = 1$. For the displaced vertex signature, we require at least one displaced vertex within $|\eta| < 2.4$ and $\epsilon = 1$. Each line corresponds to the case where a single Wilson coefficient is set to unity while all others are set to zero, with the color coding indicating the associated operator as labeled.}
    \label{fig:collidercro}
\end{figure}

The left panel of Fig.~\ref{fig:collidercro} illustrates our simulation results of   $\sigma_{\mathrm{prod}} \times \mathcal{A} \times \epsilon$   imposing the missing energy cuts described above but not including the $\mathcal{P}_D(L_1,L_2)$ factor yet. The results are evaluated at $\Lambda = 10~\mathrm{TeV}$, and each line corresponds to the prediction from only one unity Wilson coefficient, while all other coefficients are set to zero. Here the processes that generate missing–energy signals are induced by the $2 \to 3$ process, $pp \to Z'Z' j$, for the $\mathcal{O}_{1-4}$  operators, and by the $2 \to 2$ process, $pp \to Z'j$, for the $\mathcal{O}_{5,6}$ operators.  As expected, with the same couplings the production rate of $Z'$ particles via the $\mathcal{O}_{1-4}$ operators are suppressed, as the $2 \to 3$ processes, $pp \to Z' Z' j$, correspond to an additional factor of $\alpha_s$ and a high-dimensional phase space. In other words, the $\mathcal{O}_{5,6}$  operators  yield relatively larger production cross sections due to the presence of  $2 \to 2$ processes, $pp \to Z' j$.  

By adding the $\mathcal{P}_D(L_1,L_2)$ factors and imposing that the signal cross section remains below the 95\% CL upper limit \(\sigma_{\rm sig} < 0.3\,\mathrm{fb} \) \cite{ATLAS:2021kxv}, we derive the corresponding missing energy bound for a combined choice $X_1 =Y_1 =1$, shown by the light gray shaded regions in our summary plots, Fig.~\ref{fig:Zp-constraints}.  For $Z'$ particles that can be produced at LHC with little kinematical suppression and then escape from the detector, the experimental data excludes $\Lambda$ less than $3~\mathrm{TeV}$. This  is consistent with the results of searches exhibiting similar signatures within a dimension-six operator framework~\cite{Godbole:2015gma}.

\vspace{0.12cm}

Now we turn to collider search on displaced vertex signals, for $Z'$ bosons with shorter lifetimes. For such signals, the dominant $Z'$-production processes are illustrated by the first and fourth diagrams in Fig.~\ref{fig:feyndia}, that is, the $2 \to 2$ processes $gg \to Z' g$ and $gg \to Z' Z'$, which are induced by the  $\mathcal{O}_{5,6}$ and $\mathcal{O}_{1-4}$ operators, respectively. The process \( gg \to Z' g \) yields a single displaced vertex with a prompt jet, while the second one \( gg \to Z' Z' \) may produce either two displaced vertices or a single displaced vertex accompanied by missing energy. ATLAS has conducted searches for displaced vertex signals, including single displaced vertices accompanied by missing energy or by a prompt \( Z \) boson, as well as events with two displaced vertices, each containing at least two associated tracks. No significant excess over the SM prediction has been observed, leading to upper limits on the signal cross section at the level of \( \mathcal{O}(0.1~\mathrm{pb}) \) for various new physics models~\cite{ATLAS:2025pak,ATLAS:2022gbw}. We simulate such signal production processes for our model, only with one selection that requires at least one displaced vertex within $|\eta|<2.4$. 

Right panel of Fig.~\ref{fig:collidercro} shows the corresponding values of $\sigma_{\mathrm{prod}} \times \mathcal{A} \times \epsilon$ with the displaced vertex selection rule above induced by each EFT operator, again with $\epsilon = 1$ and not including the $\mathcal{P}_D(L_1,L_2)$ factor. Apparently, the dominant $Z'$-production process is $gg \to Z' Z'$  for the $\mathcal{O}_{1-4}$  operators, and $gg \to Z' g$ for the $\mathcal{O}_{5,6}$  operators. To derive the final bounds, we conservatively adopt an upper limit of $0.1\,\mathrm{pb}$ on the cross section for events with at least one displaced vertex, after applying the selection rule and the  additional factor $\mathcal{P}_D(3\,\mathrm{m},14\,\mathrm{m})$.  Again we only show the numerical results for the combination of $\mathcal{O}_{1}$  and $\mathcal{O}_{5}$ with $X_1 = Y_1 = 1$, which can be easily rescaled to other parameter choices. Being different from the missing energy case above, here one always needs non-vanishing $Y_1$ and/or $Y_2$ to make the $Z'$ particle decay. The  corresponding results are given  as the red shaded regions in the summary plot, Fig.~\ref{fig:Zp-constraints}.  
As a cross-check, we also use the partonic \(gg\!\to\!\gamma\gamma\) cross section provided in Ref.~\cite{Ellis:2018cos},  which computes from the same EFT operator to evaluate our bound. Based on the observation that their choice \(\sqrt{\hat s}=2~\mathrm{TeV}\) is a representative gluon–fusion partonic scale at the LHC, we confirm that our resulting constraints are consistent with theirs.

\subsubsection{Other Constraints at Colliders}

Even shorter-lived $Z'$ can be searched for as three-jet resonances in LHC. There have been recent studies of such for pair-produced resonances~\cite{CMS:2018ikp} and singly-produced resonances~\cite{CMS:2023tep}, thus applying to our $\mathcal{O}_{1,2,3,4}$ and $\mathcal{O}_{5,6}$ operators, separately.\footnote{Ref.~\cite{CMS:2024ldy} only looks for resonances of three quark jets, optimized for gluon jet rejections, and thus does not apply here.}  In practice, they result in  upper limits on the $Z'$ production cross section for $m_{Z'}$ between 200\,GeV and a few TeV. Lower-energy resonances of three jets cannot be resolved due to the limit jet energy resolution at LHC.  

General speaking, when the $Z'$ particle decays promptly into gluon jets, the overwhelming QCD background makes it challenging to extract strong constraints on it at LHC. For instance, using the CMS data with centre-of-mass energy $\sqrt{s} =13\,$TeV, the upper bound on the $Z'$ production cross section is tens of fb for $Z'$ masses around TeV, requiring the corresponding $\Lambda$ to be just above 1\,TeV. This scale is well below the \(pp\) centre-of-mass energy of the data and falls within (or very near) the dark-gray region of Fig.~\ref{fig:Zp-constraints}, where the EFT description breaks down at LHC energies. Therefore, we do not incorporate these bounds in our analysis. Besides, while the Large Electron-Positron (LEP) collider, among  other electron-beam experiments, are less contaminated by QCD background, their centre-of-mass energy is too low to provide competitive constraints on such high-dimensional EFT operators as well; see e.g.~\cite{Chu:2018qrm}.

In addition, the presence of new colored particles can affect the renormalization group  running of the strong coupling constant \( \alpha_s \) at high energies, thereby placing a constraint on the new physics scale \( \Lambda \gtrsim \mathcal{O}(100\,\mathrm{GeV}) \) from precision QCD measurements at   LHC~\cite{Llorente:2018wup}. Since the charge of a particle under a non-Abelian gauge group is uniquely determined, the fact that no colored new particle has been found at LHC also suggests that the new physics scale, $\Lambda $, should be above the electroweak scale. Nevertheless, a concrete bound would strongly rely on the properties of the intermediate particles, and will be deferred to our follow-up study of its UV realizations.

\subsection{Constraints from cosmological observations}

As a potentially long-lived particle, the presence of $Z^\prime$ at early Universe may leave imprints on cosmological observables. The most relevant constraints for our analysis are based on Big Bang Nucleosynthesis (BBN), the Cosmic Microwave Background (CMB), and the Late Universe observations; the last combines limits from low-redshift cosmic rays (CR) and the precisely-measured energy budget of the current Universe. These cosmological and astrophysical constraints will be studied and summarized in this subsection. Apparently, they could only constrain the parameter regions where $\tau_{Z'}$ is above $10^{-2}$\,sec, as the early Universe well before the BBN time cannot be reliably measured at this moment.  

To quantify the effects of the  $Z^\prime$  population on high-redshift Universe, we calculate its abundance via the associated Boltzmann equation: 
\begin{align}\label{eq:fullBoltzmann}
  \dot n_{Z'} + 3 H\, n_{Z'} \;=\;&
  -\,\langle\Gamma_{Z'}\rangle_{Z'  \to ggg}\!\left(n_{Z'} - n_{Z'}^{\mathrm{eq}}\right)
  \;-\; \big\langle \sigma v_{\mathrm{Møl}} \big\rangle_{Z' g \to gg}\,
        n_g^{\mathrm{eq}}\!\left(n_{Z'} - n_{Z'}^{\mathrm{eq}}\right) \notag\\
  &-\; \big\langle \sigma v_{\mathrm{Møl}} \big\rangle_{Z' Z' \to gg}\,
        \left(n_{Z'}^{2} - \big(n_{Z'}^{\mathrm{eq}}\big)^{2}\right)\,,
\end{align}
where $H$ is the Hubble expansion rate, and the other three terms on the R.H.S. stand for the decay/annihilation channels of the $Z'$ particles, together with their inverse processes. Here, \(n_{i}\) and \(n_{i}^{\mathrm{eq}}\) denote the actual number density and thermal equilibrium number density of the particle $i$, with respect to  the photon temperature $T$. The \(\sigma\) and \(v_{\mathrm{Møl}}\) denote the annihilation cross section and Møller velocity, respectively, while \(\langle ... \rangle\) indicates thermal averaging. Note that for the third process, a factor of 2, due to the fact that each process annihilate two $Z'$ particles, has canceled  with a factor of 1/2, from double-counting of events by using $n^2_{Z'}$; see Appendix~\ref{app: thermally averaged annihilation cross sections} for more details. 
Explicit expressions of the cross sections for the processes \(g Z^{\prime} \to gg\) and \(Z^{\prime} Z^{\prime} \to gg\) are shown in Appendix~\ref{app:crosssection}. The decay rate that enters the Boltzmann equation is the thermally averaged  $\langle \Gamma_{Z'} \rangle = \Gamma_{Z'}\,K_1(m_{Z'}/T)/K_2(m_{Z'}/T)$, where $K_i$ are the modified Bessel functions of the second kind~\cite{Chu:2011be}.

 \begin{figure}[t]
    \centering
    \includegraphics[width=0.495\linewidth]{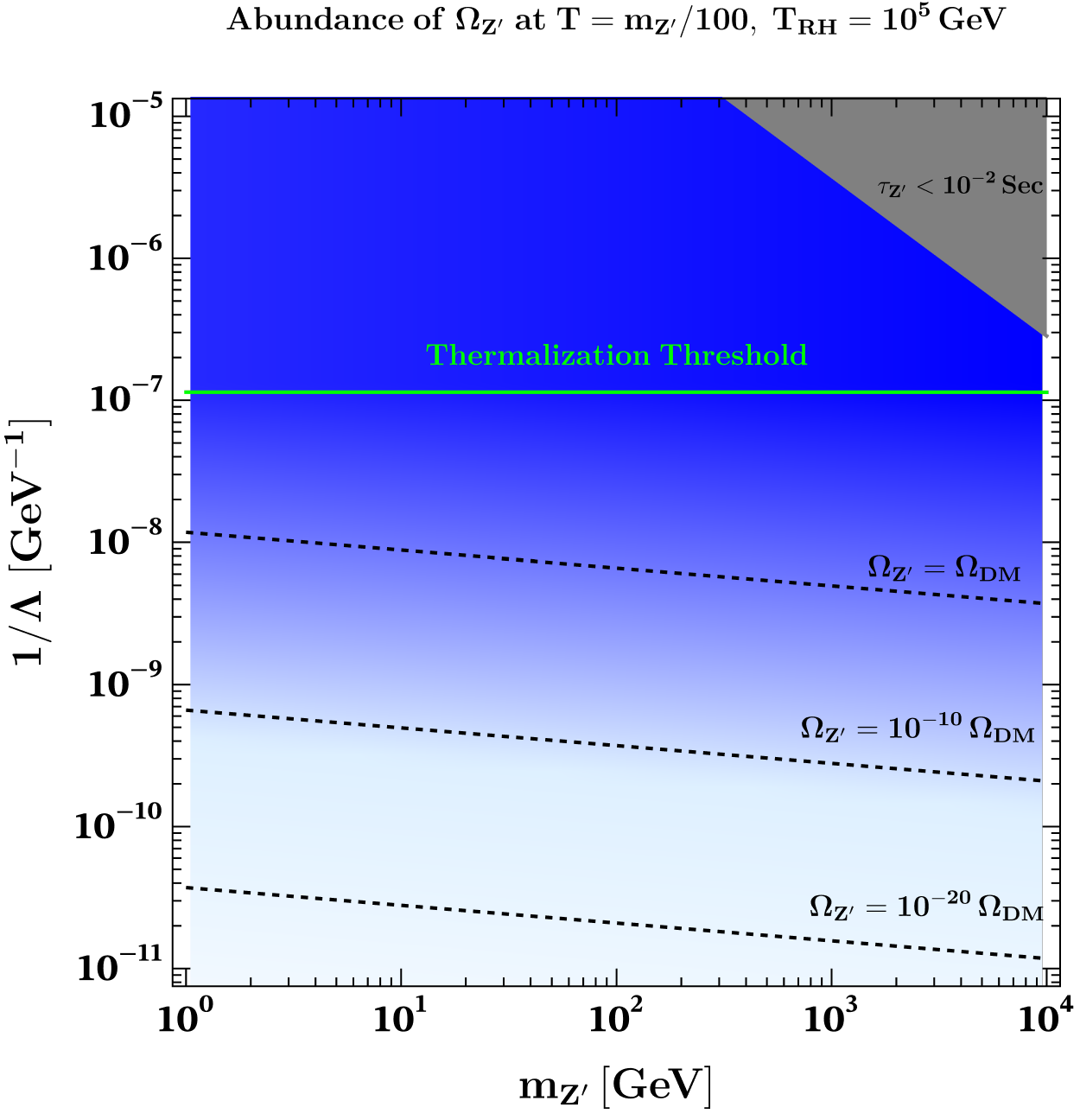}
    \includegraphics[width=0.495\linewidth]{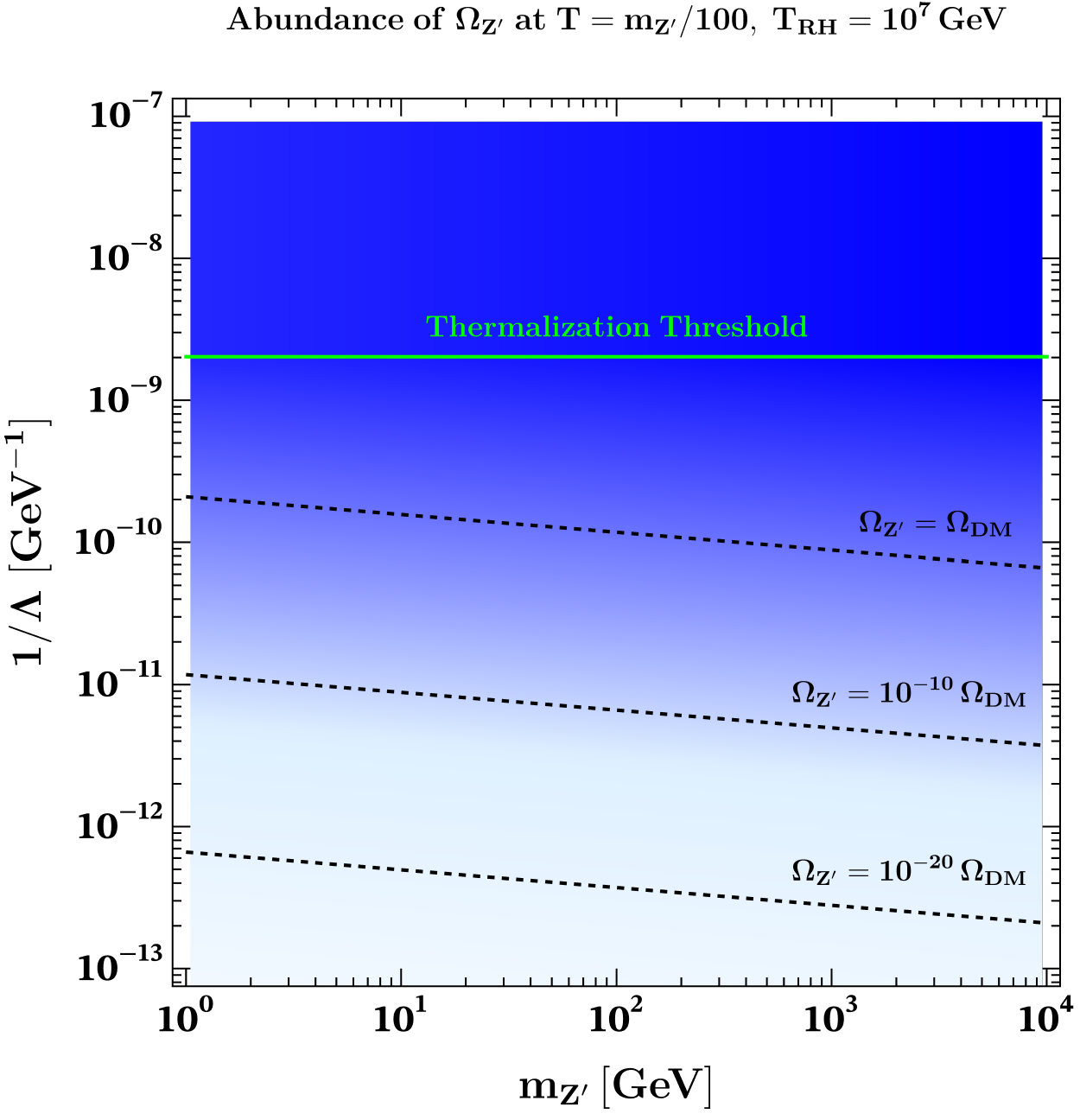}
    \caption{
Abundance of $\Omega_{Z^{\prime}}$ evaluated at $T = m_{Z^{\prime}}/100$ 
in the $(m_{Z^{\prime}}, 1/\Lambda)$ plane for 
$T_{\mathrm{RH}} = 10^{5}\,\mathrm{GeV}$ ({\bf left} panel)  and 
$10^{7}\,\mathrm{GeV}$ ({\bf right} panel), with $X_{1} = Y_{1} = 1$.
The green solid line indicates the thermalization threshold, separating the   freeze–in regime (below) from the thermalized regime (above). Three black dashed lines correspond to $\Omega_{Z'} = [1,10^{-10},10^{-20}]\,\Omega_{\mathrm{DM}}$, respectively. The dark gray region denotes $\tau_{Z'} < 10^{-2}\,\mathrm{sec}$ (only appearing in left panel due to the different selected parameter range on the y-axis),  which is irrelevant for the cosmological probes.}
    \label{fig:Ydistribution}
\end{figure}

We then solve the Boltzmann equation above for $X_1 = Y_1 = 1$ with all other operators absent. To be conservative, we take ${n}^\text{initial}_{Z^\prime} =0$ for the initial condition at the reheating temperature, $ T_\text{RH}$. Larger values of ${n}^\text{initial}_{Z^\prime} $ would result in stronger constraints if thermalization never happens. To simplify the calculation, we neglect the $Z'$ (semi-)annihilations into two gluons if both sectors are never thermalized with each other, and fix $T_{Z'}$ same as the photon temperature once  both sectors thermalize at some point. The two approximations are known to be reliable for such studies~\cite{Chu:2011be}. Our numerical calculation is then performed from the reheating temperature down to  $T = m_{Z'}/100$, where the $Z'$  production/annihilation processes necessarily decouple, and  freeze-in/out should already end.  For DM masses above $\mathcal{O}(1)$ GeV,  the age of the Universe is still within $10^{-2}\,\text{sec}$ at this moment, and $Z'$ becomes highly non-relativistic. That is, the $Z'\to ggg$ decay channel can be neglected so far. In addition,  we expect that annihilation via $Z' g \to gg$ dominates over that via $Z'Z' \to gg$ at $T\lesssim m_{Z'}$, where $n_g\gg n_{Z^\prime}$ 

The abundance $\Omega_{Z'}$ evaluated at $T = m_{Z'}/100$ in the $(m_{Z'}, 1/\Lambda)$ plane is presented in Fig.~\ref{fig:Ydistribution} for reheating temperatures of $T_{\mathrm{RH}} = 10^{5}\,\mathrm{GeV}$ (left panel) and $T_{\mathrm{RH}} = 10^{7}\,\mathrm{GeV}$ (right panel). Here we ensure the EFT validity by always restricting  $\Lambda > T_{\text{RH}}$. We also do not include the parameter space where $\tau_{Z'} < 10^{-2}\,\mathrm{sec}$, which is irrelevant for the cosmological probes considered in this work. This appears as the dark gray shaded region in left panel. As is well known, for portals that are described by EFT operators, the $Z'$ population is predominantly produced around $T \simeq T_{\mathrm{RH}} \gg m_{Z'}$, as the production processes are UV-dominated~\cite{Mambrini:2013iaa}. Additionally, its interaction strength with thermal bath increases with $T_{\mathrm{RH}}$.   Therefore one can estimate the minimal coupling needed for two-sector thermalization by comparing the interaction rate with the Hubble rate:  
\begin{equation}\label{eq:thermalCondition}
    \left\langle \sigma_{Z'g \rightarrow gg} v_{\mathrm{Møl} } \right\rangle n_{g}^{\mathrm{eq}}(T) \sim H(T)\simeq {T^2 \over M_{\mathrm{Pl}}}
\end{equation}
at $T \simeq T_{\mathrm{RH}}$, where $M_{\mathrm{Pl}}$ is the Planck mass. Recall that at relativistic limit, there exist $\sigma_{Z'g \rightarrow gg} v_{\mathrm{Møl} }  \simeq {T^6 /  \Lambda^8}$ and $n_{g}^{\mathrm{eq}}(T) \simeq T^3$, the critical value of $\Lambda$ scales as $T_{\mathrm{RH}}^{7/ 8} M_{\mathrm{Pl}}^{1/8}$, being independent of the DM mass. This is shown as the horizontal solid green lines in both panels of Fig.~\ref{fig:Ydistribution}, labeled as ``thermalization threshold''.

Below the thermalization threshold, the $Z'$ population never reaches thermal equilibrium with the SM sector, and thus its abundance at $T = m_{Z'}/100$   only increases with larger couplings.  Consequently, its frozen-in mass density, $Y_{Z'} m_{Z'}|_{T = m_{Z'}/100}$,  scales as $ \propto T_{\mathrm{RH}}^{7} m_{Z'} M_{\mathrm{Pl}}/\Lambda^{8}$, linearly dependent of mass $m_{Z'}$. For illustration, Fig.~\ref{fig:Ydistribution}  shows the three black dashed lines which correspond to $\Omega_{Z'} = [1, 10^{-10}, 10^{-20}]\,\Omega_{\mathrm{DM}}$, respectively. Based on the gray line in Fig.~\ref{fig:lifetime}, we find that the $Z^\prime$ boson is sufficiently long-lived and, through freeze-in production, can achieve the observed relic abundance, making it a viable dark matter candidate.  
 
For portal couplings above the thermalization threshold, the $Z'$ particles become able to thermalize with the gluon bath around $T = T_\text{RH}$. Note that the interaction rate, given by the L.H.S. of Eq.~\eqref{eq:thermalCondition}, decreases much faster than the Hubble rate, on the R.H.S. of the equation, when the temperature drops. As a result,  for the parameter regions we show in Fig.~\ref{fig:Ydistribution}  the decoupling of both sectors usually happens when the $Z'$ is still relativistic. That is,  the $Z'$ particle   freezes out while being still relativistic. For such relativistic freeze-out, the frozen  abundance $Y_{Z'}$ is approximately an order-one constant, and thus the resulting mass density only depends on the the mass, as $\Omega_{Z'} \propto m_{Z'}$. This explains that for parameter regions above the threshold line in Fig.~\ref{fig:Ydistribution}, the value of $\Omega_{Z'}$ becomes independent of $\Lambda$.  In fact, order-one values of  $Y_{Z'}$ with $\tau_{Z'} \ge 10^{-2}\,$sec are experimentally excluded, as will be discussed shortly below. 
 
For larger  portal couplings well above the thermalization threshold, which are not fully included in Eq.~\eqref{eq:thermalCondition} but are shown in Fig.~\ref{fig:lifetime}, the $Z'$  particle has a lifetime shorter than  $10^{-2}\,$sec. Consequently, these parameter regions are safe from existing cosmological/astrophysical probes.  At last, for even larger portal couplings, with $m_{Z'}/\Lambda \gtrsim 10^{-2}$, non-relativistic freeze-out can occur and strongly suppress the frozen $Z'$ abundance. This possibility, together with details of the freeze-in/out mechanisms, will be investigated in the next section, where the $Z'$ particle is set to be stable.

\vspace{.15cm}

Now we can study the observable consequences of an unstable $Z'$ population at $T={m_{Z'}/100}$, as given by Fig.~\ref{fig:Ydistribution}, before considering a stable $Z'$ particle in the next section. Generally speaking, its subsequent decay ejects visible energy  into the thermal bath, modifying cosmological and astrophysical observables that have been measured quite precisely.  In Fig.~\ref{fig:lifetime}, blue iso-lifetime contours demonstrate the regions of cosmological and astrophysical sensitivity: BBN ($10^{-2}\,\mathrm{sec} \lesssim \tau_{Z'} \lesssim 10^{12}\,\mathrm{sec}$) and CMB/CR ($10^{11}\,\mathrm{sec} \lesssim \tau_{Z'} \lesssim 10^{27}\,\mathrm{sec}$). Additionally,   the gray line indicates the age of the Universe, $\tau_U \simeq 4.3\times10^{17}\,\mathrm{sec}$, which serves as the criterion for whether the $Z'$ is effectively stable as a DM candidate. Since the sensitivity of each probe depends on the $Z'$ lifetime, we use complementary experimental and cosmological observations to constrain the combination of  $(m_{Z'}, 1/\Lambda)$ parameters. In what follows, we describe in detail how each of these probes constrains the parameter plane.

\begin{figure}[t]
  \centering

  \begin{subfigure}[b]{0.495\textwidth}
    \centering
    \includegraphics[width=\textwidth]{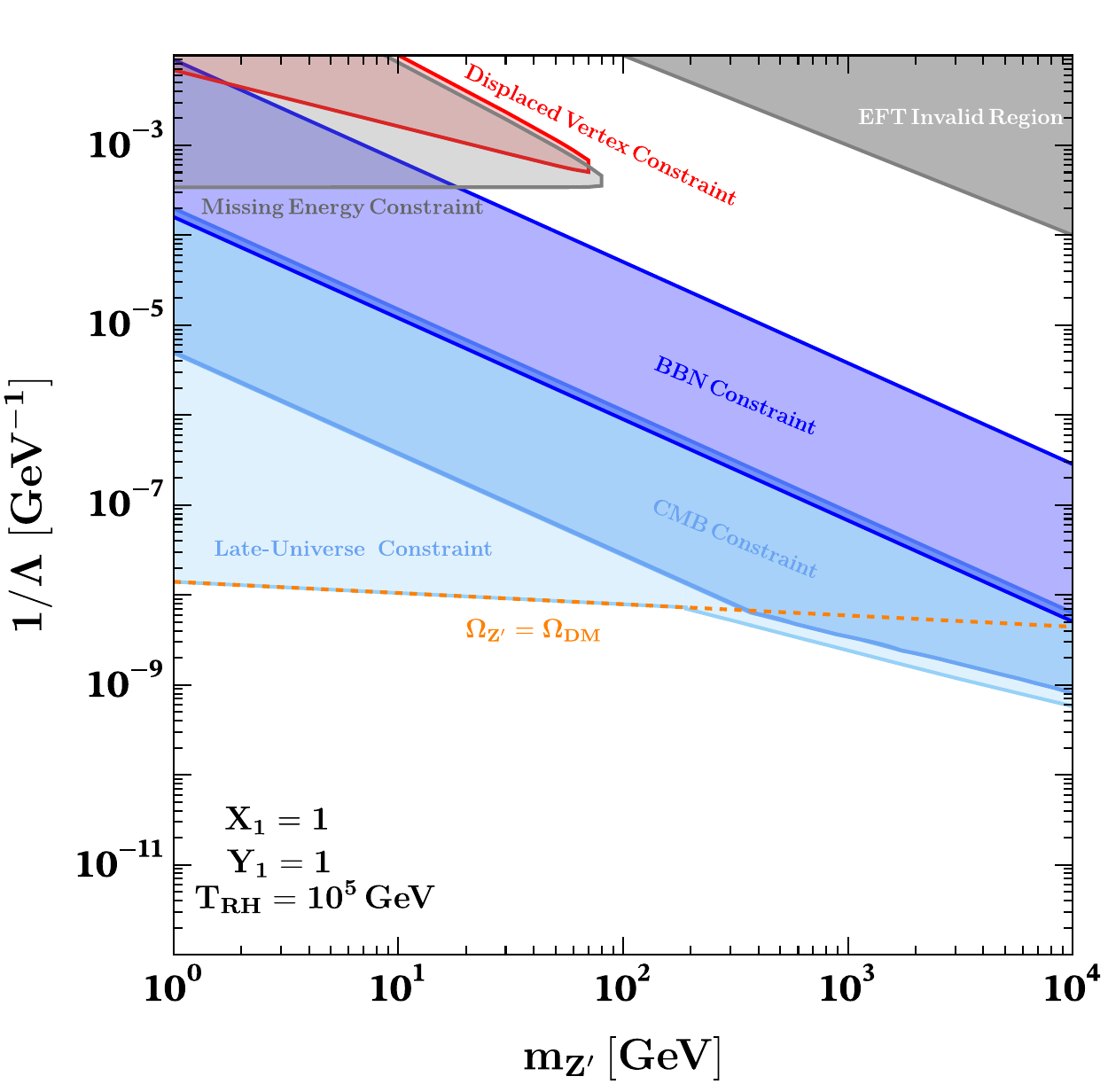}
  \end{subfigure}
  \begin{subfigure}[b]{0.495\textwidth}
    \centering
    \includegraphics[width=\textwidth]{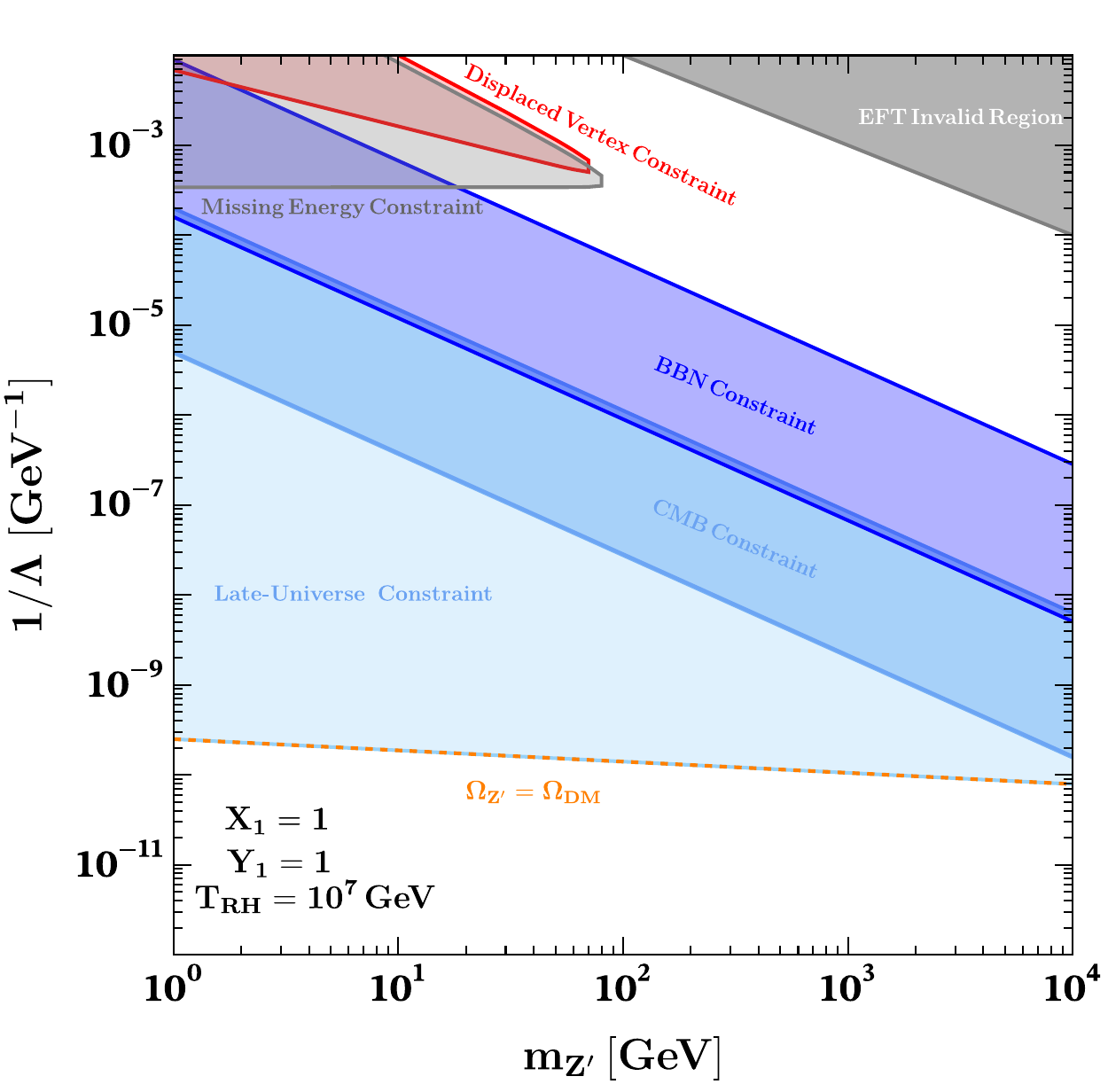}
  \end{subfigure}
\caption{Constraints on an unstable gluonic $Z'$ with $X_1 = Y_1 = 1$ with other coefficients set to zero, for reheating temperature $T_{\mathrm{RH}} = 10^5$\,GeV ({\bf left} panel) and $10^7$\,GeV ({\bf right} panel). In both panels, collider bounds from missing-energy searches~\cite{ATLAS:2021kxv} are indicated by the light-gray regions, while displaced-vertex searches~\cite{ATLAS:2025pak,ATLAS:2022gbw} exclude the red regions. Cosmological limits from BBN~\cite{Kawasaki:2004qu,Angel:2025dkw},   CMB~\cite{Acharya:2019uba}, and late-Universe observations~\cite{Planck:2018vyg,Paopiamsap:2023uuo,Fermi-LAT:2012pls} are shown as distinct blue shaded regions.  Orange dashed lines indicates $\Omega_{Z'} = \Omega_{\rm DM}$ generated via the freeze-in mechanism. The dark-gray area marks the EFT-invalid region, where $\Lambda \le m_{Z'}$, while the EFT approach may already become invalid for collider constraints at $1/\Lambda  \gtrsim 10^{-3}\,\text{GeV}^{-1}$; same for Fig.~\ref{fig:Zp1-constraints} below.}
  \label{fig:Zp-constraints}
\end{figure}

\emph{BBN.} The BBN theory successfully  predicts the primordial abundances of light elements, while late-decaying \( Z' \) particles may disrupt these predictions through hadronic energy injections that alter the \( n/p \) ratio or induce hadro-dissociation of light nuclei. Recent searches use primordial abundance measurements of deuterium, helium-3/4, and lithium-7 to place strong constraints on the primordial abundance of BSM particles decaying into the SM bath, with lifetimes between \( \tau = 10^{-2} \, \mathrm{sec} \) and \( 10^{12} \, \mathrm{sec} \) \cite{Kawasaki:2004qu,Angel:2025dkw}. Therefore, we adopt their results to  constrain the \((m_{Z'},1/\Lambda)\) plane by requiring the primordial \(Z'\) abundance calculated above to lie below the observational upper bounds. 

\emph{CMB.} The injection of high–energy electromagnetic particles around recombination modifies the recombination history and hence the CMB anisotropy power spectra. Accounting for delayed energy deposition, there are many studies that place stringent limits on the  abundance of late–decaying BSM particles with lifetimes \(\tau \sim 10^{11}\text{–}10^{24}\,\mathrm{sec}\). In our setup, \(Z'\) decays predominantly via \(Z'\!\to ggg\); the ensuing hadronization rapidly yields photons and charged leptons (e.g., via \(\pi^0\!\to\gamma\gamma\)). Accordingly, we assume that most of the energy injected by \(Z'\) decay is electromagnetic and, for each \(\tau_{Z'}\), require the \(Z'\) relic abundance to satisfy the corresponding CMB bounds in Ref.~\cite{Acharya:2019uba}, thereby constraining the \((m_{Z'},\,1/\Lambda)\) plane. 

\emph{Late Universe (CR + relic abundance).} For lifetimes exceeding the age of the Universe ($\tau_{Z'}>\tau_U$), the $Z'$ is effectively stable on cosmological timescales and can constitute a DM component. In this case, CR data constrain the injected hadronic power from dark–matter decay. Under the conservative assumption of purely hadronic decays, CR observations imply an approximately mass–independent lower bound on the lifetime for single–component DM, \(\tau_{\rm DM}\gtrsim 10^{27}\,\mathrm{sec}\), over the mass range relevant this work~\cite{Paopiamsap:2023uuo,Fermi-LAT:2012pls}. Since the constrained quantity is the injected power \(P_{\rm inj}\propto \Omega_{Z'}/\tau_{Z'}\) in the limit of $\tau_{Z'} \gg \tau_U$, such observations can also be re-scaled to constrain a sub-leading component of the DM abundance via 
\begin{equation}
    \tau_{Z'}\;\gtrsim\; \Omega_{Z'}/\Omega_{\rm DM} \times 10^{27}\,\mathrm{sec}\,.
\end{equation}
In addition, we impose the relic-density bound $\Omega_{Z'}h^2\le\Omega_{\rm DM}h^2\simeq0.12$, as measured by Planck~\cite{Planck:2018vyg}.  We refer to the CR and relic abundance bounds collectively as the late–Universe constraint. 

The resulting exclusion limits are displayed as distinct blue shaded regions, corresponding to the BBN, CMB and late-Universe probes, in the left (right) panel of Fig.~\ref{fig:Zp-constraints} for $T_{\mathrm{RH}} = 10^{5}\,\mathrm{GeV}$ ($T_{\mathrm{RH}} = 10^{7}\,\mathrm{GeV}$), respectively. Note that, in deriving the cosmological constraints, we discard the part of the cosmologically relevant parameter space where the EFT condition $\Lambda \gtrsim T_{\mathrm{RH}}$ is violated, since our EFT approach does not provide reliable predictions there. However, the collider bounds obtained in Sec.~\ref{sec:collider} remain reliable, as they probe much lower energy scales where the EFT description can be mostly considered as  valid.

In both panels of Fig.~\ref{fig:Zp-constraints}, the dashed orange line denotes the single–component DM condition, $\Omega_{Z'} = \Omega_{\mathrm{DM}}$. Its physical meaning is clarified in Fig.~\ref{fig:Ydistribution}, where this line separates the region with $\Omega_{Z'} > \Omega_{\mathrm{DM}}$ (above the line) from that with $\Omega_{Z'} < \Omega_{\mathrm{DM}}$ (below the line) at the time where $T = m_{Z'}/100$. In the left panel of Fig.~\ref{fig:Zp-constraints} with $T_\text{RH} =10^5\,$GeV, we infer that the $Z'$ can account for the entirety of the DM abundance only for $m_{Z'} \lesssim 200\,\mathrm{GeV}$. Above this mass, indirect searches~\cite{Paopiamsap:2023uuo,Fermi-LAT:2012pls} require $\Omega_{Z'} < \Omega_{\mathrm{DM}}$, implying that the $Z'$ can only constitute a subdominant fraction of the DM abundance, otherwise its decay would lead to gamma-ray excess beyond the current experimental limit. Similarly, CMB data~\cite{Acharya:2019uba} imposes  upper limits on the primordial $Z'$ abundance that intersects the single–component DM contour at $\tau_{Z'} \sim 10^{24}\,\mathrm{sec}$. For shorter lifetimes, these constraints become increasingly stringent, forcing $\Omega_{Z'} < \Omega_{\mathrm{DM}}$ until $\tau_{Z'} \simeq 10^{12}\,\mathrm{sec}$. Complementarily,  BBN data~\cite{Kawasaki:2004qu,Angel:2025dkw} provide the leading constraints for $Z'$ particles with lifetime between  $10^{-2}$ and $10^{12}$\,sec, requiring $\Omega_{Z'} < \Omega_{\mathrm{DM}}$ in this region. This is the reason why, above the dotted orange line of $\Omega_{Z'} = \Omega_{\mathrm{DM}}$ line, the combination of BBN and CMB exclude all the parameter regions for lifetime  \(10^{-2}\,\mathrm{sec}\lesssim\tau_{Z'}\lesssim10^{12}\,\mathrm{sec}\) and \(10^{12}\,\mathrm{sec}\lesssim\tau_{Z'}\lesssim10^{24}\,\mathrm{sec}\),  as indicated in Fig.~\ref{fig:lifetime}. 

At last, we emphasize that for the line of $\Omega_{Z'} = \Omega_{\mathrm{DM}}$ in Fig.~\ref{fig:Zp-constraints}, the observed DM relic abundance can be produced via pair annihilation of SM gluons. Its location relies on the reheating temperature. 
For the left panel with  $T_\text{RH} = 10^5\,$GeV, the lifetime of $Z'$ is too short to satisfy the current CMB and CR bounds if $m_{Z^\prime} >200$\,GeV, but the region for lighter $Z'$  is not constrained by observations, except for the argument that its current density should not exceed the observed DM density. That is, $Z'$ can play the role of the dominant DM candidate, and  a successful production via the freeze-in mechanism is allowed. It happens for  $m_{Z'} \le 200\,$GeV in the case of   $T_\text{RH} = 10^5\,$GeV, and applies to the whole ${Z'}$ mass region considered here if  $T_\text{RH} = 10^7\,$GeV, as shown in the right panel of Fig.~\ref{fig:Zp-constraints}. 
This is because, in the freeze-in mechanism, higher reheating temperatures correspond to smaller couplings, which result in longer-lived $Z'$ particles. These longer lifetimes help alleviate the CMB and CR bounds, thereby allowing for a wider ${Z'}$ mass range to account for the whole observed DM abundance.

\section{Gluonic $Z'$ as a Dark Matter Candidate}
\label{sec.dm}
In this section, we further investigate the potential of such a gluonic $Z'$ as the leading DM component and the consequent signatures. We have studied this model above under the assumption that the Wilson coefficients \(X_i\) for $\mathcal{O}_{1-4}$ and \(Y_i\) for $\mathcal{O}_{5,6}$ are of the same order, where the cosmological stability requirement for dark-matter candidate, \(\tau_{Z'}\gtrsim 10^{17}\,\mathrm{sec}\), implies its extremely weak couplings to the SM sector, as shown as the solid gray line in Fig.~\ref{fig:lifetime}.  Consequently, it severely limits the parameter region, as well as the possibility to probe this DM model non-gravitationally. For this reason, below we impose a \(\mathcal{Z}_2\) symmetry for the $Z'$ particle, which forbids its decay  via $\mathcal{O}_{5,6}$ operators. As discussed earlier, such a residual \(\mathcal{Z}_2\) symmetry can arise if $Z'$ is originally belong to a non-abelian gauge group in the dark sector. This in  turn allows us to consider two production mechanisms of the observed DM relic abundance: freeze-in and freeze-out, as well as the associated DM phenomenology, such as direct/indirect detections, and collider searches.

\subsection{Freeze-in Production}\label{sec.freezein}
\begin{figure}[t]
\centering
  \begin{subfigure}[b]{0.495\textwidth}
    \centering
    \includegraphics[width=\textwidth]{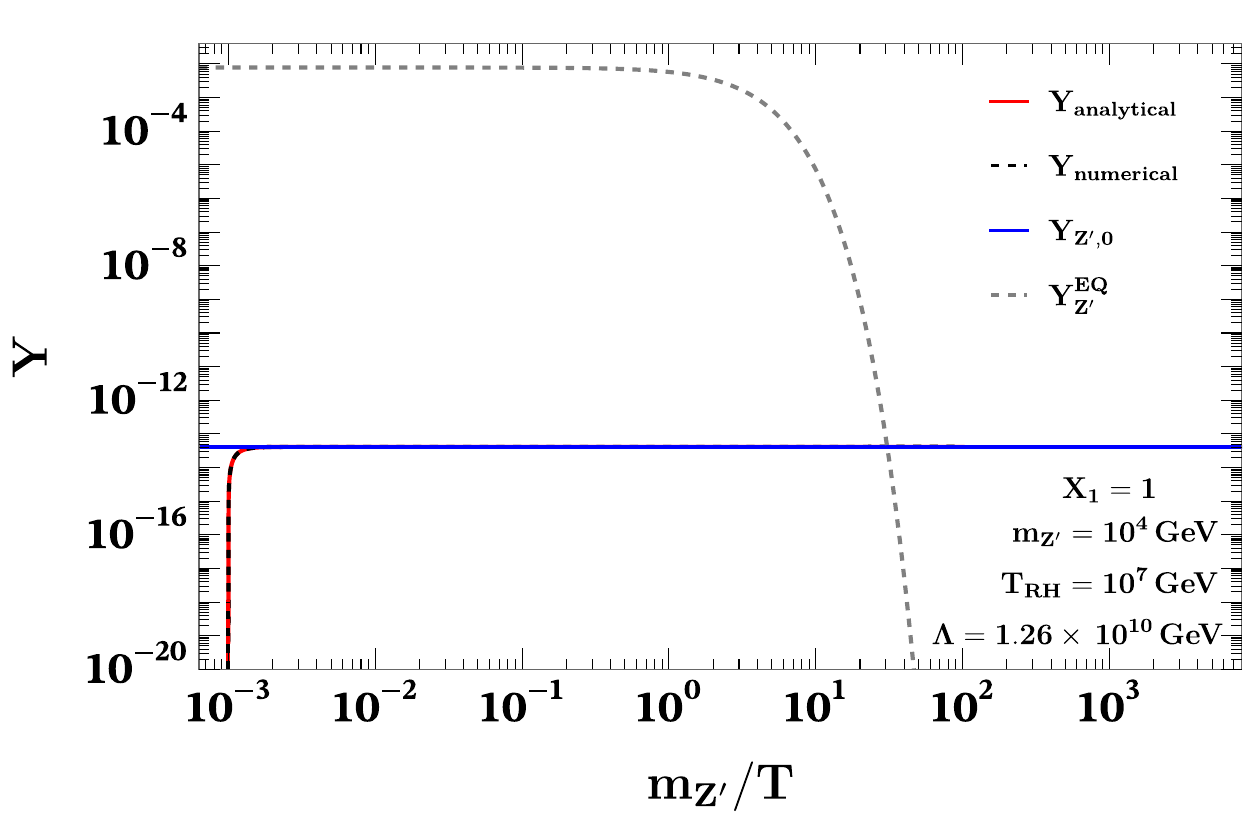}
  \end{subfigure}
  \begin{subfigure}[b]{0.495\textwidth}
    \centering
    \includegraphics[width=\textwidth]{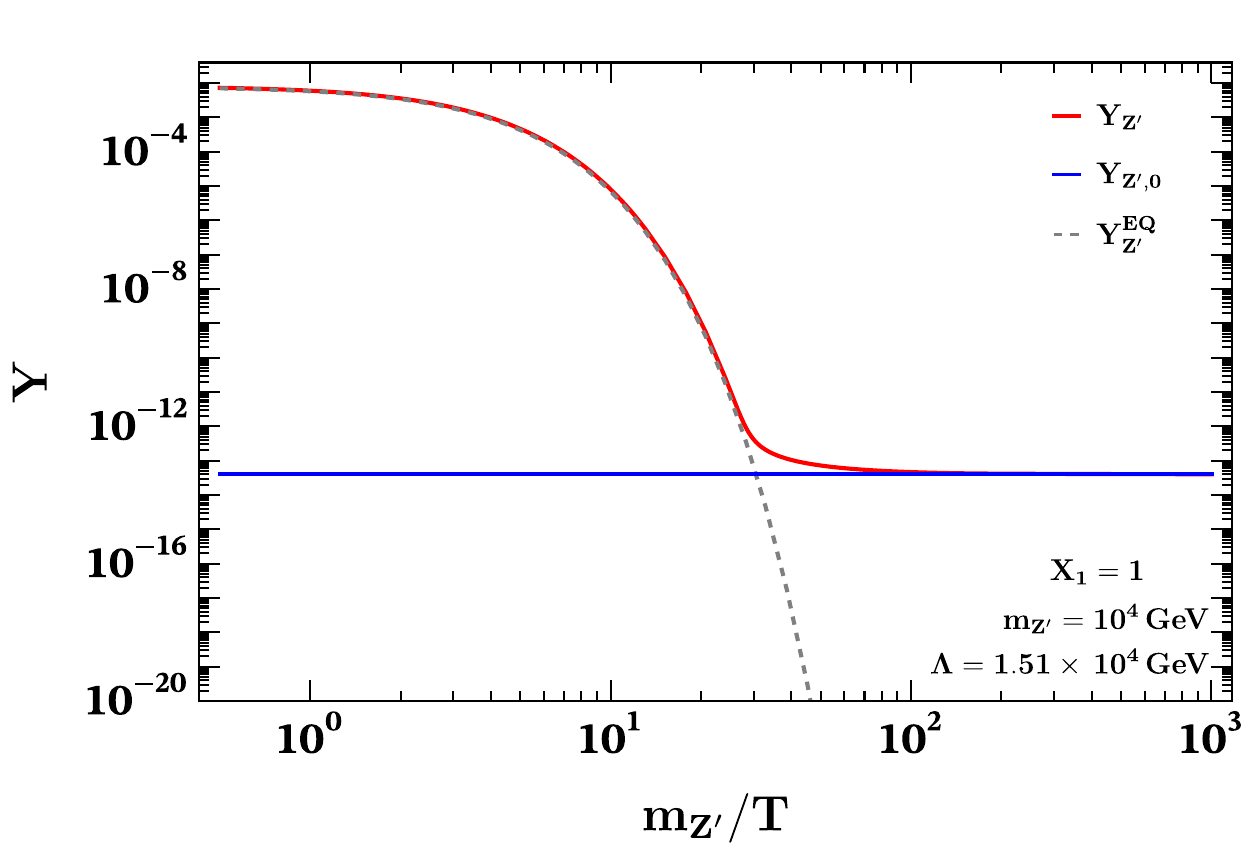}
  \end{subfigure}
    \caption{{\bf Left} panel: Evolution of the \( Z' \)   abundance in the freeze-in scenario, evaluated for \( \Lambda = 1.26 \times 10^{10}~\mathrm{GeV} \), \( m_{Z'} = 10^4~\mathrm{GeV} \), \( T_{\mathrm{RH}} = 10^7~\mathrm{GeV} \), and \( X_1 = 1 \). The analytical prediction (red solid line) is compared to the numerical solution (black dashed line), while the expected final yield \( Y_{Z',0}= 4.1 \times 10^{-14}\) for this benchmark scenario, is indicated by the horizontal blue line. The gray dashed line indicates the corresponding thermal equilibrium abundance, $Y_{Z'}^{\rm EQ}$. {\bf Right} panel: Evolution of the \( Z' \) relic abundance in the freeze-out scenario for the benchmark point with \( X_1=1 \) and all other Wilson coefficients set to zero, taking \( \Lambda=1.51\times10^4\,\mathrm{GeV} \) and \( m_{Z'}=10^4\,\mathrm{GeV} \). The red solid line shows the time-dependent abundance \( Y_{Z'} \), the gray dashed line denotes its corresponding thermal equilibrium value \( Y_{Z'}^{\mathrm{EQ}} \), and the horizontal blue line indicates the observed relic abundance at present, \( Y_{Z',0} \). 
}\label{fig:freezein}
\end{figure}

In the freeze-in mechanism~\cite{McDonald:2001vt,Choi:2005vq,Kusenko:2006rh,Petraki:2007gq,Hall:2009bx}, SM particles are produced directly from the inflaton decay and form a thermal bath first, while the DM particles, the \( Z' \) bosons in our model,  reside in a dark sector and are initially absent at the end of the inflation. Through feeble portal interactions induced by the operators $\mathcal{O}_{1-4}$, the population of \( Z' \) particles are later generated via out-of-equilibrium processes from the SM bath, such as  gluon fusion
\begin{equation}
    g(p_1) + g(p_2) \rightarrow Z'(k_1) + Z'(k_2)\,,
\end{equation}
eventually yielding the observed relic abundance. As a special case of Eq.~\eqref{eq:fullBoltzmann}, the corresponding Boltzmann equation that describes the freeze-in evolution of  \( n_{Z'} \) can be obtained by omitting contributions of $\mathcal{O}_{5,6}$ and  inverse processes as follows: 
\begin{equation}\label{eq.Boltzmann_equation}
    \dot{n}_{Z'} + 3 H(T) n_{Z'} = R(T) \equiv      \big\langle \sigma v_{\mathrm{Møl}}  \big\rangle_{ gg \to Z' Z'}\big(n_{g}^{\mathrm{eq}}\big)^{2} \,,
\end{equation}
where  $R(T)$ denotes the $Z'$ production rate. More explicit expressions for these quantities are provided in Appendix~\ref{app:freezein}. In contrast to IR-dominated freeze-in scenario, where the DM production proceeds through renormalizable operators and is insensitive to the reheating temperature \( T_{\text{RH}} \), our setup involves \( Z' \) production via non-renormalizable dimension-8 operators, corresponding to a UV-dominated freeze-in mechanism; see~\cite{Bernal:2017kxu} and references therein. In this case, DM particles are predominantly produced just after reheating, rendering the final abundance highly sensitive to \( T_{\text{RH}} \). In order to make the EFT description valid, we always require that $T_{RH}$ is much smaller than the cut-off energy scale $\Lambda$. The present-day frozen-in yield, defined as $Y_{Z',0} \equiv n_{Z',0}/s_0$, where $n_{Z',0}$ is the present-day number density of $Z'$ and $s_0$ is the present-day entropy density, thus admits an analytical expression:
\begin{equation}\label{eq.abundance_general}
    Y_{Z',0} \simeq \frac{11153.2\, F_1(\{X_j\})\, M_{\mathrm{Pl}}\, T_{\mathrm{RH}}^7}{\Lambda^8\, \pi^7\, g_*^S(T_{\mathrm{RH}})\, \sqrt{g_*^\rho(T_{\mathrm{RH}})}}\,,
\end{equation}
where  $g_*^{S}(T)$ and $g_*^{\rho}(T)$ denote the effective numbers of relativistic degrees of freedom for entropy and energy densities, respectively. The dimensionless function \( F_1(\{X_j\}) \) encodes the dependence on the Wilson coefficients \( X_j \). The derivation of Eq.~\eqref{eq.abundance_general} and the explicit form of \( F_1(\{X_j\}) \) are provided in Appendix~\ref{app:freezein}. 
For numerical results, the yield can be approximated as
\begin{equation}\label{eq.abundance_numeric}
Y_{Z',0} \simeq 2.61 \times 10^{-13}  \left( \frac{T_{\mathrm{RH}}}{10^7\,\mathrm{GeV}} \right)^7 \left( \frac{F_1(\{X_j\})}{64} \right) \left( \frac{10^{10}\,\mathrm{GeV}}{\Lambda} \right)^8.
\end{equation}
The left panel of Fig.~\ref{fig:freezein} illustrates the evolution of \(Y_{Z'}\) during the freeze-in process for a benchmark scenario in which only \(X_1 = 1\) contributes (i.e., \(F_1(\{X_j\}) = 64\)), with \(\Lambda = 1.26 \times 10^{10}~\mathrm{GeV}\), \(m_{Z'} = 10^4~\mathrm{GeV}\), and \(T_{\mathrm{RH}} = 10^7~\mathrm{GeV}\). The expected yield, \(Y_{Z',0} = 4.1 \times 10^{-14}\), for this benchmark scenario, computed according to Eq.~\eqref{eq.abundance_numeric}, is indicated by the horizontal blue line. For comparison, the gray dashed line indicates the corresponding thermal equilibrium abundance, $Y_{Z'}^{\rm EQ}$. The evolution obtained from numerical integration and from the analytical approximation is represented by the black dashed and red solid lines, respectively. The excellent agreement between these two results confirms the validity of the analytical expression, and the final yield is consistent with the observed DM relic abundance.

Then, by requiring that the \( Z' \) population saturates the observed DM relic abundance, \( \Omega_{Z',0} h^2 \simeq 0.12 \), corresponding to \(m_{Z'}\,Y_{Z',0} \simeq  4.1 \times 10^{-10}\,\text{GeV}\), as measured by the Planck Collaboration~\cite{Planck:2018vyg}, we derive the corresponding values of \(1/\Lambda\) as a function of the mediator mass \(m_{Z'}\). For illustration, we consider two benchmark reheating temperatures, \( T_{\mathrm{RH}} = 10^5~\mathrm{GeV} \) and \( 10^7~\mathrm{GeV} \), and assume that only the Wilson coefficient \( X_1 = 1 \) contributes (i.e., \( F_1(\{X_j\}) = 64 \)). These are shown in right panel of Fig.~\ref{fig:Zp1-constraints} by solid (dashed) orange lines for \( T_{\mathrm{RH}} = 10^5\)~(\( 10^7\))~GeV. In both cases, the resulting \(1/\Lambda\) values remain within the regime of EFT validity, i.e., \( \Lambda > T_{\mathrm{RH}} \).

\subsection{Freeze-out Production}\label{sec.freezeout}
The freeze-out mechanism is a widely studied framework for explaining the DM relic abundance; see e.g. ~\cite{Scherrer:1985zt, Kolb:1990vq, Gondolo:1990dk}. Unlike the freeze-in mechanism above, where the observed DM abundance is produced out-of-equilibrium, the freeze-out mechanism assumes that DM was initially in equilibrium. As the Universe expanded and cooled, DM particles eventually decoupled from the thermal bath, freezing their comoving number density. With a residual $\mathcal {Z}_2$ symmetry, the number density of \(Z^\prime\) is governed by the Boltzmann equation (Eq.~\ref{eq:fullBoltzmann}) through the \(Z^\prime Z^\prime \leftrightarrow g g\) channel:
\begin{equation}
    \dot{n}_{Z^\prime} + 3 H n_{Z^\prime} =  -  \langle \sigma v_{\mathrm{Møl}} \rangle_{Z' Z' \to gg} \left( n_{Z'}^{2} - \left( n_{Z'}^{\mathrm{eq}} \right)^{2} \right) \,,
\end{equation}
where the thermally-averaged cross section \(\langle \sigma v_{\mathrm{Møl}} \rangle_{Z' Z' \to gg}\) can be calculated from~\cite{Gondolo:1990dk}:
\begin{equation}\label{eq:freezeoutR}
   \langle \sigma v_{\mathrm{Møl}} \rangle_{Z' Z' \to gg} = \frac{1}{8 m_{Z^\prime}^4 T K_2^2({m_\text{Z'} \over T})} \int_{4 m_{Z^\prime}^2}^{\infty}(s - 4 m_{Z^\prime}^2) \sqrt{s} \, K_1({\sqrt{s} \over T}) \, \sigma_{Z' Z' \to gg} \, \mathrm{d}s\,,
\end{equation}
where $T$ is the photon temperature and $K_{1,2}$ are modified Bessel functions of the second kind. The explicit form of $\sigma_{Z' Z' \to gg}$ is given in Appendix~\ref{app:crosssection}.

We numerically compute the thermal evolution of the \( Z' \) relic abundance and require that its present value matches the observed DM abundance \( \Omega_{Z',0} h^2 \simeq 0.12 \), within the benchmark scenario where only \( X_1 = 1 \) contributes. The result is shown in right panel of Fig.~\ref{fig:freezein}  for \( m_{Z'}=10^4\,\mathrm{GeV} \) and  \( \Lambda=1.51\times10^4\,\mathrm{GeV} \). In the plot, the red solid line represents the time-evolved relic abundance $Y_{Z^\prime}$, the red dashed line corresponds to its thermal equilibrium abundance $Y_{Z\prime}^{EQ} (T)$, and the horizontal blue solid line indicates the observed DM abundance $Y_{Z',0}$.  The general freeze-out annihilation cross section, 
\(\langle \sigma v_{\mathrm{M\o l}} \rangle_{Z' Z' \to gg}\),  required to reproduce the observed DM relic abundance, as a function of $m_{Z'}$ is illustrated as the dotted red line in the left panel of Fig.~\ref{fig:Zp1-constraints}. The obtained values are close to 
the typically canonical one, \(10^{-36}\,\mathrm{cm}^2\). In turn, the couplings needed to generate the observed DM relic abundance, as a function of the DM mass $m_{Z'}$, is shown as the solid red line in the right panel of Fig.~\ref{fig:Zp1-constraints}.

\subsection{Direct detection}

Direct detection of DM relies on observing stable nuclear (electron) recoils resulting from elastic DM--nucleus\,(electron) scattering processes. Thus the null results from recent direct detection experiments~\cite{LZ:2024zvo,PandaX:2022xqx,DarkSide-20k:2024yfq} place upper limits on the spin-independent DM--nucleon scattering cross-section. For $Z^\prime$ with \(m_{Z'}\gtrsim 1\, \mathrm{GeV}\) and a typical Galactic speed \(v\sim10^{-3}c\), \(Z'\)--nucleon scattering is non-relativistic, with the maximal momentum transfer,  $q = 2\,\mu_{Z'n}\,v$, set by the reduced mass,
\begin{equation}
  \mu_{Z'n}\equiv \frac{m_{Z'}m_n}{m_{Z'}+m_n}\,,
\end{equation}
where \(m_n\simeq 0.94~\mathrm{GeV}\) denotes the nucleon mass, and the factor of 2 corresponds to backward scattering (maximal nucleon recoil).

\begin{figure}[t]
  \centering

  \begin{subfigure}{0.495\textwidth}
    \centering
    \includegraphics[width=\textwidth]{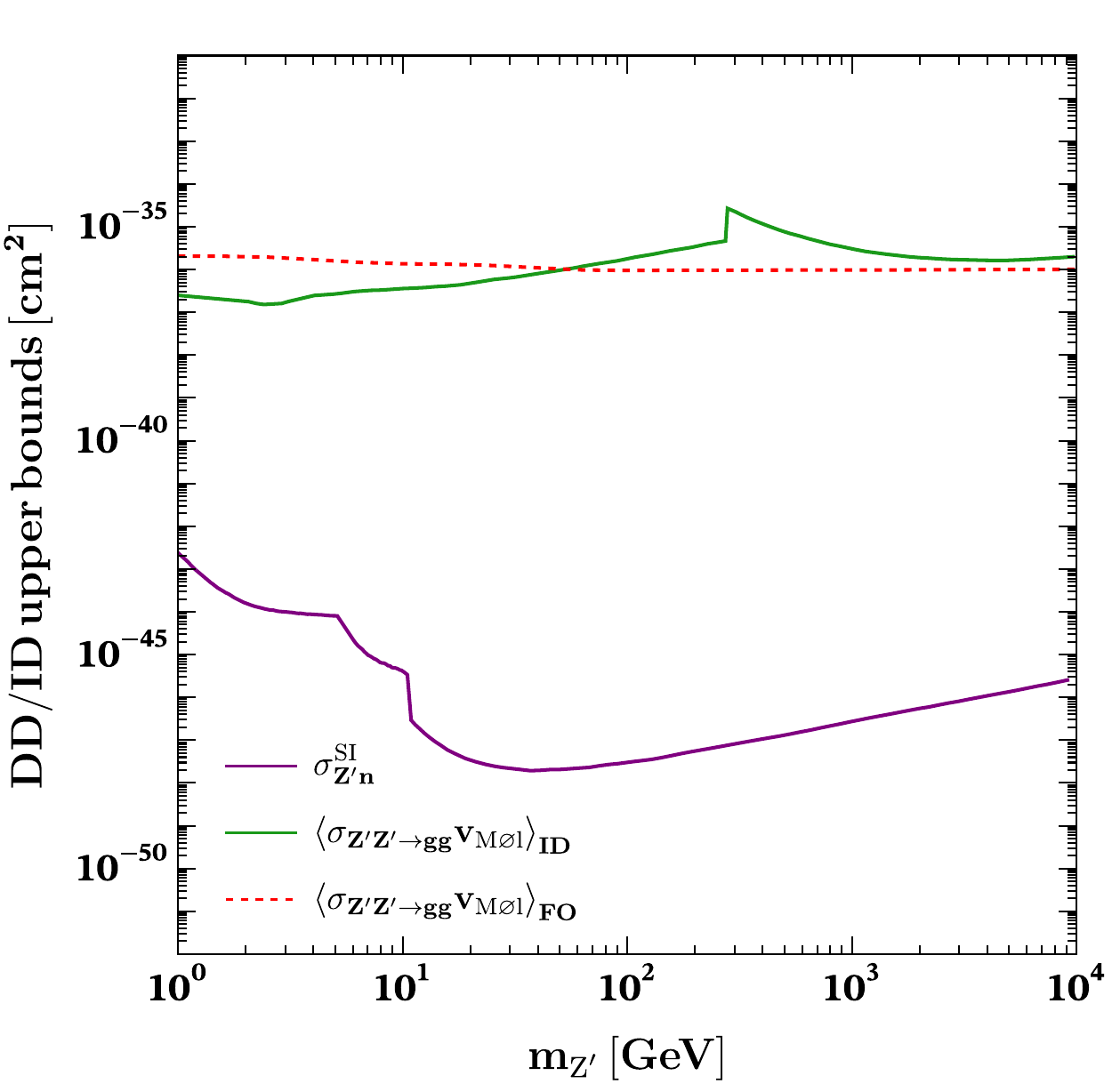}
  \end{subfigure}
  \begin{subfigure}{0.495\textwidth}
    \centering
    \includegraphics[width=\textwidth]{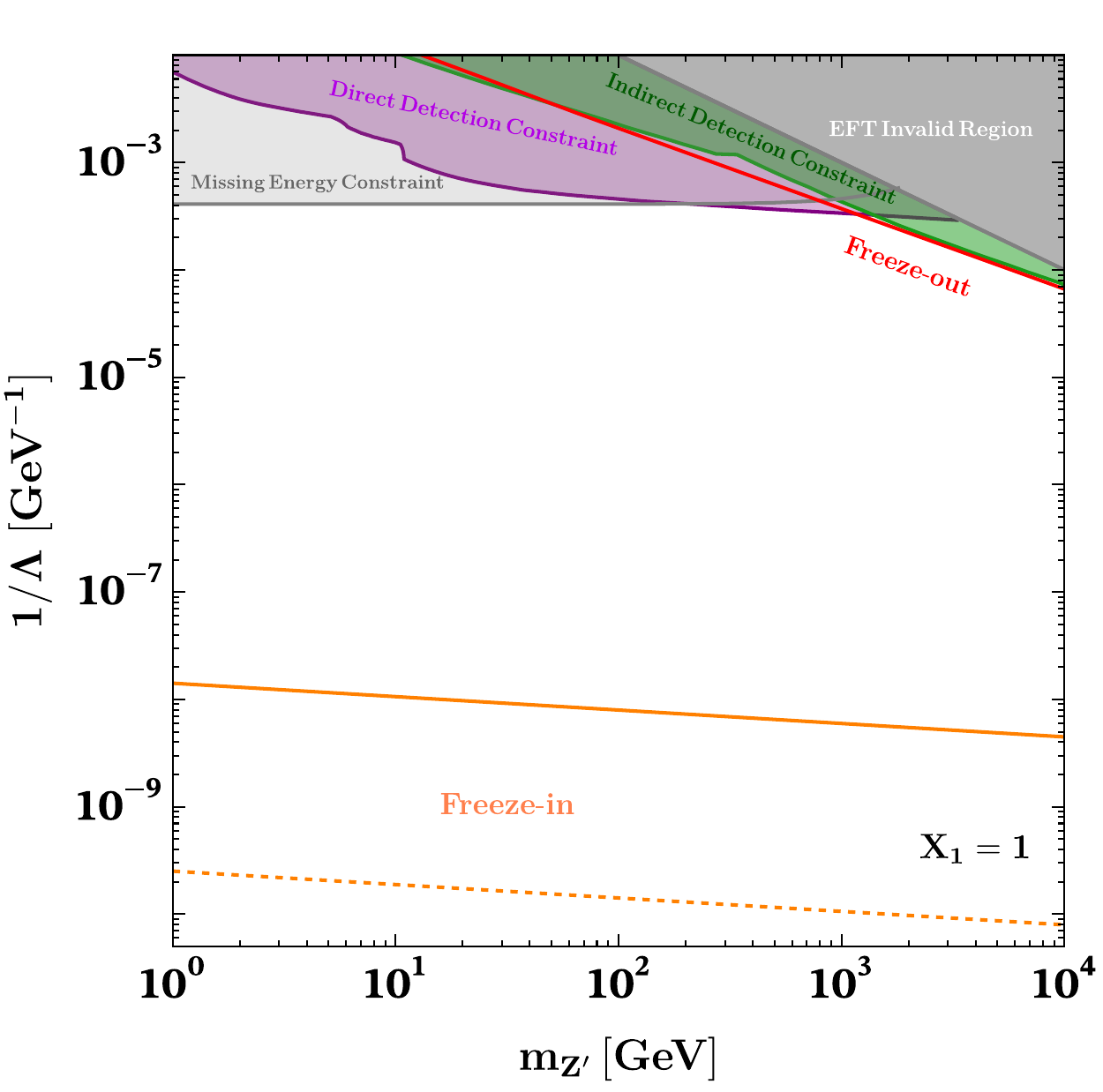}
  \end{subfigure}

\caption{{\bf Left} panel shows  upper bounds  adopted here on DM-nucleon scattering  cross sections (purple solid line) by combining LZ~\cite{LZ:2024zvo}, PandaX-4T~\cite{PandaX:2022xqx}, and DarkSide-50~\cite{DarkSide-20k:2024yfq} results, as well as upper bounds on DM gluonic annihilation cross sections (green solid line), which combine \textsc{H.E.S.S.}~\cite{HESS:2022ygk} and \textit{Fermi}-LAT~\cite{Paopiamsap:2023uuo} data. Dotted red line indicates the annihilation cross sections needed for the observed DM abundance in the freeze-out mechanism at non-relativistic limit.  
{\bf Right} panel gives bounds on $1/ \Lambda$ for \(Z'\) as a DM candidate, assuming that only \(X_1 = 1\) while all other coefficients are set to zero. The light-gray, purple, and green shaded regions are excluded by missing-energy~\cite{ATLAS:2021kxv}, direct-detection~\cite{LZ:2024zvo,PandaX:2022xqx,DarkSide-20k:2024yfq}, and indirect-detection~\cite{HESS:2022ygk,Paopiamsap:2023uuo} searches, respectively. The red line gives the correct DM relic abundance via thermal freeze-out, while the solid (dashed) orange lines indicates the parameters required to produce the correct DM relic abundance via freeze-in  for reheating temperatures \(T_{\mathrm{RH}} = 10^5~\mathrm{GeV}\) (\(10^7~\mathrm{GeV}\)).}
  \label{fig:Zp1-constraints}
\end{figure}

Since \(q\ll\Lambda_{\rm QCD}\),  the interaction must be described in terms of nucleon matrix elements rather than partonic perturbation theory. Accordingly, computing the spin-independent cross section \(\sigma^{\rm SI}_{Z'N}\) requires matching the partonic operators \(\mathcal O_{1\text{--}4}\) onto nucleon-level operators. Recent studies on such matching procedures can be found in Refs.~\cite{Fitzpatrick:2012ix,Bishara:2016hek,Bishara:2017pfq}, though only the matching of the partonic operator \( \mathcal{O}_1 \) has been thoroughly studied in these works. Generally, according to power counting, the other operators contribute to the same order of magnitude as \( \mathcal{O}_1 \).  So we first consider the case with \( X_1 = 1 \) and all other Wilson coefficients set to zero, as illustration. In the non-relativistic limit, the spin-independent $Z'$--nucleon scattering cross-section is
\begin{equation}
  \sigma_{Z' n}^{\rm SI} = \frac{256\pi\, m_G^2\, m_n^2\, m_{Z'}^4}{81\alpha_s^2\Lambda^8(m_n + m_{Z'})^2}\,,
  \label{eq:sigmaZn}
\end{equation}
where \( m_G \simeq 848~\mathrm{MeV} \) denotes the gluonic contribution to the nucleon mass in the isospin-symmetric limit and \( \alpha_s \simeq 0.118 \) is the QCD coupling constant \cite{Bishara:2017pfq}.  The detailed derivation of $\sigma_{Z' n}^{\rm SI}$ can be found in Appendix~\ref{app:crosssection}.
We combine the latest results from LZ~\cite{LZ:2024zvo}, PandaX-4T~\cite{PandaX:2022xqx}, and DarkSide-50~\cite{DarkSide-20k:2024yfq} to obtain  upper bounds on \(\sigma_{Z'n}^{\rm SI}\), as demonstrated by  the purple solid line in the left panel of Fig.~\ref{fig:Zp1-constraints}. The corresponding direct-detection constraints on $1/\Lambda$, obtained by requiring $\sigma_{Z'n}^{\mathrm{SI}}$ to lie below this bound, are shown as purple regions in the right panel of Fig.~\ref{fig:Zp1-constraints}. These results exclude the thermal freeze-out mechanism for $m_{Z'} \lesssim 1~\text{TeV}$.

Recasting precisely the constraints for the other operators is non-trivial. Instead, we compute the relevant partonic-level cross sections  (mostly importantly, \( Z' g \to Z' g \)) induced by the operators $\mathcal{O}_{2-4}$, and find that their contributions are of the same order as that from \( \mathcal{O}_1 \) at non-relativistic limit. For simplicity, we therefore assume that the \( Z' N \to Z' N \) scattering cross sections induced by $\mathcal{O}_{2-4}$ are of the same order as that by \( \mathcal{O}_1 \). Based on this,  we believe that direct detection bounds on $\Lambda$ for $\mathcal{O}_{2-4}$ are comparable, due to the high exponent on $\Lambda$  in the scattering cross sections. 

~~\\

\subsection{Indirect detection}
While  $Z'$ pair annihilation becomes highly suppressed with the expansion of the Universe, and has a negligible impact on the relic density, its annihilation at high-redshifts and local dense regions can still product a lot of gluons, or neutral/charged pions after hadronization. This would potentially lead to excesses of gamma-ray, anti-proton, and high-energy neutrinos, beyond the predictions of standard astrophysics; see e.g.~\cite{Bertone:2004pz, Cirelli:2010xx, Chu:2012qy, IceCube:2022clp}.
 Therefore, observational data from gamma-ray and cosmic-ray observatories, such as the High Energy Stereoscopic System (H.E.S.S.) \cite{Hinton:2004eu}, Alpha Magnetic Spectrometer (AMS-02)~\cite{AMS:2016oqu}, and the Fermi-LAT telescope~\cite{Fermi-LAT:2009ihh}, can be used to place upper limits  on the DM annihilation cross section at various masses. 
 
For illustration, we again adopt the parameter choice where \( X_1 = 1 \), while setting all other Wilson coefficients to vanish. In this case the  DM annihilation cross section at  the non-relativistic limit can be written as 
\begin{equation}\label{eq.annihilationcross}
  \left\langle \sigma_{Z^\prime Z^\prime\rightarrow gg} v_{\mathrm{Møl}} \right\rangle = \frac{32\sqrt{2} \, m_{Z^\prime}^6}{3\pi \Lambda^8}\,.
\end{equation}
We combine the latest bounds on the upper bounds on \(\left\langle \sigma_{Z^\prime Z^\prime \rightarrow gg} v_{\mathrm{M\o l}} \right\rangle\) from the \textsc{H.E.S.S.}~\cite{HESS:2022ygk} and \textit{Fermi}-LAT~\cite{Paopiamsap:2023uuo} collaborations, as given by the green solid line in the left panel of Fig.~\ref{fig:Zp1-constraints}. The corresponding exclusions on our $1/\Lambda$ are then given as the green regions at the top-right corner of the right panel of Fig.~\ref{fig:Zp1-constraints}. For DM annihilation dominantly into gluons, thermal freeze-out is only experimentally allowed   for the $Z'$ mass above tens of GeV, in agreement with previous studies~\cite{Elor:2015bho, Bartlett:2022ztj}.

\subsection{Collider search and other probes}
In the DM scenario, the $Z'$ can be produced at the LHC but gives rise only to  missing–energy signatures through the $2 \to 3$ process $pp \to Z' Z' j$. In contrast to the general case discussed in Sec.~\ref{sec.me_dv}, we here omit the probability factor $\mathcal{P}_D(L_1,L_2)$, and impose the DM $Z'$ does not decay inside the detector. That is, the relevant signal cross section   is simply 
\begin{equation}
    \sigma_{\mathrm{sig}} = \sigma_{\mathrm{prod}} \times \mathcal{A} \times \epsilon\,,
\end{equation}
whose numerical values has already been given in the left panel of Fig.~\ref{fig:collidercro} as a function of $m_{Z'}$. Considering the case $X_1 = 1$, we impose the same requirement $\sigma_{\rm sig} < 0.3~\mathrm{fb}$~\cite{ATLAS:2021kxv} as before, and   obtain the  missing–energy constraints as the light gray shaded regions in the right panel of Fig.~\ref{fig:Zp1-constraints}.

Before ending this section, we briefly comment on other potential probes of such a DM abundance, such as its capture in stars and consequent effects. For instance,  in a DM-dense environment, the processes $Z' Z' \to gg$ and $Z' g \to Z' g$ in a neutron–star core could in principle modify the stellar  evolution. For the small effective couplings (large $\Lambda$) considered here , the corresponding rates are parametrically suppressed due to the low centre-of-mass energy, and are not expected to yield competitive constraints, so we do not further consider them here.

\section{Conclusions}
\label{sec:conclu}
A new   gauge boson \(Z'\) that couples to the SM predominantly through gluons—via heavy mediators charged under both dark and \(SU(3)_C\) gauge symmetries—provides a well-motivated target for a rich phenomenological study due to its longevity. In this work we construct the corresponding EFT operators by integrating out the mediators, thereby generating the gluon-portal interaction. Two benchmark scenarios are considered: (i) \(Z'\) is unstable and can decay, and (ii) an additional \(\mathcal{Z}_2\) symmetry forbids its decay, rendering \(Z'\) strictly stable.

The scenario in which the \( Z' \) is allowed to decay can be searched for at LHC, via its signatures as either monojet plus missing energy or displaced vertex. We have derived stringent constraints for \( m_{Z'} \lesssim 100~\mathrm{GeV} \) by requiring that the predicted missing energy and displaced vertex production cross section does not exceed the upper limits set by recent ATLAS searches. Moreover,  we constrain the parameter space of this scenario using energy-injection limits from the BBN data,   CMB measurements, and late-Universe probes, covering the $Z'$ lifetimes in the range of $10^{-2}\text{–}10^{27}\,\mathrm{sec}$. The  exclusions derived from these collider and cosmological/astrophysical experiments are summarized in Fig.~\ref{fig:Zp-constraints}, for different reheating temperatures. Our results suggest that the LHC data mostly probe the new physics scale $\Lambda$ below $10^4$\,GeV in this scenario. In contrast, the combination of BBN, CMB and late-Universe observations can be sensitive to   $\Lambda$ values as high as  $10^9$\,GeV, as long as the $Z'$ particle has a lifetime longer than $10^{-2}$ \,sec.

For the other scenario, where the \(Z'\) particle is stabilized by a \(\mathcal{Z}_2\) symmetry, it is cosmologically stable and can serve as a DM  candidate in both freeze-in/out production mechanisms. Under the assumption that $Z'$ dominates the observed DM abundance, we constrain the parameter space of this scenario by using the latest results from direct detection experiments,  such as LZ, PandaX-4T and DarkSide-50, as well as  from indirect detection  ones, H.E.S.S. and Fermi-LAT. The  combined constraints on \(Z'\)  are shown in  Fig.~\ref{fig:Zp1-constraints}. In addition, we give the portal couplings needed to  generate the observed DM relic abundance via both freeze-in and freeze-out mechanisms. As is expected, for portal interactions induced by EFT operators (or equivalently, very heavy mediators), the freeze-in mechanism remains well beyond the experimental sensitivity at present, while the combination of direct/indirect detection and collider experiments are close to fully exclude the thermal freeze-out mechanism for DM masses below a few TeV.

One key novelty of this framework is the longevity of the $Z'$ particle.  Under very simple assumptions on its UV completion, $Z'$  can be shown to only couple to the SM particles via at least dimension-8 operators, and thus is naturally long-lived. In fact, even without imposing a \(\mathcal{Z}_2\) symmetry to make it  absolutely stable, our $Z'$ particle, as the dominant DM component, can accommodate the freeze-in  mechanism  for $m_{Z'} \lesssim 10^2\text{-} 10^4$\,GeV, without violating any experimental bounds. This is in sharp contrast with the general kinetic mixing models, where  imposing the freeze-in mechanism and indirect detection constraints would require a dark photon DM to have masses well below the two-electron-mass threshold~\cite{Delaunay:2020vdb, Caputo:2021eaa}. We leave a full construction of the underlying UV completion, together with the possible EFT operators such UV completion can generate, for future work~\cite{our:followup}.

\begin{acknowledgments}
X. C. is supported by the National Natural Science Foundation of China (grant No.\,E4146602), and the Fundamental Research Funds for the Central Universities (grant No.\,E4EQ6605X2 and E5ER6601A2).  T.M. is partly supported by Chinese Academy of Sciences Pioneer Initiative "Talent Introduction Plan" (grant No.\,E4ER6601A2), the Fundamental Research Funds for the Central Universities (grant No.\,E4EQ6602X2), and the National Natural Science Foundation of China (grant No.\,E514660101). H.L. is supported by the U.S. Department of Energy under Grant Contract~DE-SC0012704 and would like to acknowledge the hospitality of the ICTP-AP, where this work was initiated.

\end{acknowledgments}

\appendix

\section{The Decay Width of $Z^\prime \to ggg$}\label{app.decaycalc}

We begin the Appendix by calculating the decay width for the process
\begin{equation}
    Z^\prime(p) \rightarrow g(k_1) + g(k_2) + g(k_3)
\end{equation}
 using the leading-order interaction induced by the operators $\mathcal{O}_{5,6}$, as defined in Eq.~(\ref{eq:ZGGG}),
\begin{equation}
\mathcal{L}_{Z' g g g}
= \frac{Y_1}{4\Lambda^4}\, d_{abc}\,
\partial_{[\nu} Z'_{\alpha]} \,
\partial_{[\alpha} G^a_{\beta]} \,
\partial_{[\beta} G^b_{\mu]} \,
\partial_{[\mu} G^c_{\nu]}
+ \frac{Y_2}{4\Lambda^4}\, d_{abc}\,
\partial_{[\alpha} Z'_{\beta]} \,
\partial_{[\alpha} G^a_{\beta]} \,
\partial_{[\mu} G^b_{\nu]} \,
\partial_{[\mu} G^c_{\nu]} \,,
\end{equation}
where the symmetric structure $d_{abc}$ arises from the color trace of the generators associated with $\mathcal{O}_{5,6}$,
\begin{equation}
\mathrm{Tr}\!\left[T^a T^b T^c\right]
= \frac{1}{4}\left(d^{abc} + i f^{abc}\right)\,,
\end{equation}
and the antisymmetric structure constants $f^{abc}$ do not contribute to the interaction, as they vanish upon contraction with the symmetric Lorentz structures of the operators $\mathcal{O}_5$ and $\mathcal{O}_6$.  In the Lie algebra of $SU(3)_C$, $d_{abc}$ ($f_{abc}$) denotes the totally symmetric (antisymmetric) invariant tensor defined by the anticommutator (commutator) of the fundamental generators $T_a$,
\[
\{T_a,T_b\} = \frac{1}{3}\delta_{ab}\,\mathbf{1} + d_{abc}\,T_c\,,~~~
[T_a,T_b] = if_{abc}\,T_c\,,
\]
which encode the symmetric (anti-symmetric) part of the product of generators.  The color factor $\mathcal{C}$ for the squared amplitude can be computed as
\begin{equation}
\mathcal{C}=\frac{1}{16}|d_{abc}|^2 = \frac{5}{6}\,.
\end{equation}
The color-stripped amplitude square after summing over the polarization is 
\begin{equation}\label{eq.amplitudey}
\begin{aligned}
\sum_{\rm pol}|\mathcal{M}|^2 &=\frac{8}{ \Lambda^8}  \bigg[ 
\left( 13 Y_1^2 + 56 Y_1 Y_2 + 80 Y_2^2 \right) \big( (k_1 \cdot p)^2 (k_2 \cdot k_3)^2 \big)  \\
&\quad +\quad \left( 13 Y_1^2 + 56 Y_1 Y_2 + 80 Y_2^2 \right) \big( (k_1 \cdot k_3)^2 (k_2 \cdot p)^2 \big) \\
&\quad + \left( 13 Y_1^2 + 56 Y_1 Y_2 + 80 Y_2^2 \right) \big( (k_1 \cdot k_2)^2 (k_3 \cdot p)^2 \big) \\
&\quad - 4 (Y_1 + 4 Y_2)^2 (k_1 \cdot p) (k_2 \cdot k_3) \Big[ (k_1 \cdot k_3) (k_2 \cdot p) + (k_1 \cdot k_2) (k_3 \cdot p) \Big] \\
&\quad - (Y_1 + 4 Y_2)^2 (k_1 \cdot k_2) (k_1 \cdot k_3) \Big[ 4 (k_2 \cdot p) (k_3 \cdot p) - 3 p^2 (k_2 \cdot k_3) \Big] \bigg]\,.
\end{aligned}
\end{equation}
The spin- and color-averaged amplitude squared is expressed as
\begin{equation}
\overline{|\mathcal{M}|^2} = \frac{1}{3} \sum_{\rm pol.} \sum_{\rm col.} |\mathcal{M}|^2 = \frac{\mathcal{C}}{3} \sum_{\rm pol.} |\mathcal{M}|^2\,,
\end{equation}
where the factor of ${1 /3}$ is the $Z^\prime$ spin-averaging factor. Using Lorentz invariant variables $s_{12} \equiv (k_1 + k_2)^2$ and $s_{23} \equiv (k_2 + k_3)^2$, the decay width $\Gamma_{Z^{\prime} \rightarrow g g g}$  is given by
\begin{equation}
\Gamma_{Z^{\prime} \rightarrow g g g} = \frac{1}{\mathcal{S}} \frac{1}{256\pi^3 m_{Z'}^3} \int_{s_{12}^{\rm min}}^{s_{12}^{\rm max}} \! \! \! \int_{s_{23}^{\rm min}(s_{12})}^{s_{23}^{\rm max}(s_{12})} \overline{|\mathcal{M}|^2}\, \mathrm{d}s_{23} \, \mathrm{d}s_{12}\,,
\end{equation}
where $\mathcal{S}=3 !$ is the final state symmetry factor, and the phase space integral limit is :
\begin{equation}
\begin{aligned}
&s_{12}\in [0,~m_{Z'}^2]\,,\\
&s_{23}\in  [0,~m_{Z'}^2 - s_{12}]\,.
\end{aligned}
\end{equation}
In the end, the total decay width that combines the contributions $\mathcal{O}_5$ and $\mathcal{O}_6$ is thus:
\begin{equation}
\Gamma_{Z^\prime \to ggg} = \frac{m_{Z'}^9}{41472 \pi^3 \Lambda^8} (2 Y_1^2 + 7 Y_1 Y_2 + 8 Y_2^2)\,.
\end{equation}

\section{Calculation of relevant cross-sections{}}\label{app:crosssection}

In the section, we provide   detailed calculations of  relevant $2$-body cross sections used in the main text, including those for both $Z'$ creation/annihilation and $Z'$--nucleon scattering.  

\subsection{\texorpdfstring{\( gZ^\prime\to gg  \)}{} process}

Firstly, we consider the process
\begin{equation}
    Z^\prime(p) + g(k_3) \rightarrow g(k_1) + g(k_2)\,.
\end{equation}
According to the crossing symmetry, the color-stripped amplitude squared is identical to Eq.~(\ref{eq.amplitudey}). Using the Mandelstam variable $t = (p - k_1)^2 = (k_3 - k_2)^2$, the color-stripped amplitude squared is given by:
\begin{equation}
    \begin{aligned}
\sum_{\rm pol}\left|\mathcal{M}\right|^2 =& \frac{1}{2\Lambda^8} \bigg[ -2 s \left(-m_{Z^\prime}^2 + s + t\right) \left(m_{Z^\prime}^2 (-2 s + t) + 2 s (s + t)\right) (Y_1 + 4 Y_2)^2 \\
& \quad -4 \left(m_{Z^\prime}^2 - t\right) t \left(-2 s^2 - 2 s t - t^2 + m_{Z^\prime}^2 (2 s + t)\right) (Y_1 + 4 Y_2)^2 \\
& \quad + \left(m_{Z^\prime}^2 - s\right)^2 s^2 \left(13 Y_1^2 + 56 Y_1 Y_2 + 80 Y_2^2\right) \\
& \quad + \left(m_{Z^\prime}^2 - t\right)^2 t^2 \left(13 Y_1^2 + 56 Y_1 Y_2 + 80 Y_2^2\right) \\
& \quad + (s + t)^2 \left(-m_{Z^\prime}^2 + s + t\right)^2 \left(13 Y_1^2 + 56 Y_1 Y_2 + 80 Y_2^2\right) \bigg].
\end{aligned}
\end{equation}
The integration limits for the Mandelstam variable \( t \) are determined by the kinematic constraints:
\begin{equation}
\label{eq:integrallimit}
t_{\text{min/max}}
= \left[ E_{\text{cm}}(s, m_1, m_2) - E_{\text{cm}}(s, m_3, m_4) \right]^2
  - \left[ P_{\text{cm}}(s, m_1, m_2) \pm P_{\text{cm}}(s, m_3, m_4) \right]^2,
\end{equation}
where the center-of-mass energies and momenta are defined as:
\begin{equation}
E_{\text{cm}}(s, m_a, m_b) = \frac{s + m_a^2 - m_b^2}{2 \sqrt{s}}\,, \qquad 
P_{\text{cm}}(s, m_a, m_b) = \sqrt{E_{\text{cm}}^2(s, m_a, m_b) - m_a^2}\,.
\end{equation}
Finally, the unpolarized cross section for the inverse process $Z^\prime g \rightarrow gg$ is given by:
\begin{equation}\label{eq.zgggcrosssection}
    \begin{aligned}
        \sigma_{Z^\prime g\rightarrow gg} &= \frac{1}{2!}\frac{1}{3\times16}\frac{1}{64\pi s} \cdot \frac{1}{P_{\text{cm}}(s, m_{Z^\prime}, 0)^2}\sum_{\rm pol}\sum_{\rm col.}\int_{t_{\min}}^{t_{\max}} \left|\mathcal{M}\right|^2 \mathrm{d}t \\
        &= \frac{s-m_{Z^\prime}^2 }{55296\pi\Lambda^8} \Big[
            4 m_{Z^\prime}^2 s (7 Y_1^2 + 20 Y_1 Y_2 + 16 Y_2^2) \\
            &\quad + m_{Z^\prime}^4 (11 Y_1^2 + 40 Y_1 Y_2 + 48 Y_2^2) \\
            &\quad + 21 s^2 (11 Y_1^2 + 40 Y_1 Y_2 + 48 Y_2^2)
        \Big]\,.
    \end{aligned}
\end{equation}
\subsection{ \texorpdfstring{\( Z'Z' \to gg \)}{} process}
In this subsection,  we consider the process 
\begin{equation}
    Z^\prime(p_1)+Z^\prime(p_2) \rightarrow g(k_1) + g(k_2)\,.
\end{equation}
At the leading order with operators $\mathcal{O}_{1-4}$, the Lagrangian can be expressed as  
\begin{equation}
\begin{aligned}
\mathcal{L}{Z'Z'gg} &= \frac{X_1}{2\Lambda^4} \delta^{ab} \partial_{[\alpha} Z^\prime_{\beta]} \partial_{[\alpha} Z^\prime_{\beta]} \partial_{[\mu} G^a_{\nu]} \partial_{[\mu} G^{b}_{\rho]} \
+ \frac{X_2}{2\Lambda^4} \delta^{ab}  \partial_{[\mu} Z^\prime_{\beta]} \partial_{[\alpha} Z^\prime_{\nu]} \partial_{[\alpha} G^a_{\nu]} \partial_{[\mu} G^{b}_{\beta]} \
\\&+ \frac{X_3}{2\Lambda^4} \delta^{ab}  \partial_{[\nu} Z^\prime_{\beta]} \partial_{[\alpha} Z^\prime_{\nu]} \partial_{[\alpha} G^a_{\mu]} \partial_{[\mu} G^{b}_{\beta]} \
+ \frac{X_4}{2\Lambda^4} \delta^{ab}  \partial_{[\mu} Z^\prime_{\beta]} \partial_{[\nu} Z^\prime_{\alpha]} \partial_{[\alpha} G^{a}_{\mu]} \partial_{[\beta} G^{b}_{\nu]}\,. 
\end{aligned}
\end{equation}
The color-stripped amplitude squared after summing over the polarization is
\begin{equation}\label{eq.amplitudex}
\begin{aligned}
\sum_{\rm pol}|\mathcal{M}|^2
= \frac{16}{ \Lambda^8} \Bigg\{ &
(\vec{k}_1 \cdot \vec{k}_2)^2 \bigg[
  2 A \, (\vec{p}_1 \cdot \vec{p}_2)^2
+ M_v^4 B
\bigg]
\\[4pt]
&+ 2 \bigg[
  C\, (\vec{k}_1 \cdot \vec{p}_2)^2 (\vec{k}_2 \cdot \vec{p}_1)^2
+ C\, (\vec{k}_1 \cdot \vec{p}_1)^2 (\vec{k}_2 \cdot \vec{p}_2)^2
\\
&\hspace{3em}
- 4 (2X_2 + X_4)(4X_1 + X_3 - X_4)
  (\vec{k}_1 \cdot \vec{p}_1) (\vec{k}_1 \cdot \vec{p}_2)
  (\vec{k}_2 \cdot \vec{p}_1) (\vec{k}_2 \cdot \vec{p}_2)
\bigg]
\\[4pt]
&+ 2 (2X_2 + X_4) (\vec{k}_1 \cdot \vec{k}_2) \bigg[
  (\vec{k}_1 \cdot \vec{p}_2) \Big(
    M_v^2 (\vec{k}_2 \cdot \vec{p}_2)\, D
  - 2 E\, (\vec{k}_2 \cdot \vec{p}_1)(\vec{p}_1 \cdot \vec{p}_2)
  \Big)
\\
&\hspace{7.2em}
+ (\vec{k}_1 \cdot \vec{p}_1) \Big(
    M_v^2 (\vec{k}_2 \cdot \vec{p}_1)\, D
  - 2 E\, (\vec{k}_2 \cdot \vec{p}_2)(\vec{p}_1 \cdot \vec{p}_2)
  \Big)
\bigg]
\Bigg\}\,,
\end{aligned}
\end{equation}
where the coefficients \( A \), \( B \), \( C \), \( D \), and \( E \) are defined as
\begin{equation}
\begin{aligned}
A &\equiv 32X_1^2 + 8X_1(2X_2 + 2X_3 + X_4)
     + 4X_2^2 + 4X_2X_3 + 2X_3^2 + 2X_3X_4 + X_4^2\,, \\[2pt]
B &\equiv 32X_1^2 + 16X_1X_3 + 4X_2X_3 + 8X_2X_4 + 3X_3^2 + 2X_3X_4\,, \\[2pt]
C &\equiv 8X_1(2X_2 + X_4) + 12X_2^2 + 8X_2(X_3 + X_4)
     + X_3^2 + 4X_3X_4 + X_4^2\,, \\[2pt]
D &\equiv 8X_1 + 2X_2 + 2X_3 - X_4\,, \\[2pt]
E &\equiv 4X_1 + 2X_2 + X_3\,.
\end{aligned}
\end{equation}
Using the Mandelstam variable $t = (p - k_1)^2 = (k_3 - k_2)^2$ and the integration limits defined in Eq.~(\ref{eq:integrallimit}), 
the unpolarized cross section for the process $Z' Z' \rightarrow g g$ is
\begin{equation}\label{eq.zzggcrosssection}
\begin{aligned}
\sigma_{Z' Z' \to g g}(s)
&= \frac{1}{2!}\,\frac{1}{9}\,\frac{1}{64\pi s}\,
    \frac{1}{P_{\rm cm}(s, m_{Z'}, m_{Z'})^2}
    \sum_{\rm pol.} \sum_{\rm col.}
    \int_{t_{\min}}^{t_{\max}} \left|\mathcal{M}\right|^2 \, {\rm d} t
\\
&= \frac{s^{3/2}}{1080 \pi\, \Lambda^8 \sqrt{s - 4 m_{Z'}^2}}
   \Big[ 2 m_{Z'}^4 \,\mathcal{A}
       - 2 m_{Z'}^2 s \,\mathcal{B}
       + s^2 \,\mathcal{C} \Big] \,,
\end{aligned}
\end{equation}
where we have defined the coefficient
\begin{equation}
\begin{aligned}
\mathcal{A} &\equiv
  1440 X_1^2
+ 240 X_1 (2 X_2 + 3 X_3 + X_4)
+ 116 X_2^2
+ 192 X_2 X_3
+ 116 X_2 X_4
+ 108 X_3^2
+ 96 X_3 X_4
+ 29 X_4^2 \,,
\\[2pt]
\mathcal{B} &\equiv
  960 X_1^2
+ 160 X_1 (2 X_2 + 3 X_3 + X_4)
+ 88 X_2^2
+ 116 X_2 X_3
+ 8 X_2 X_4
+ 69 X_3^2
+ 58 X_3 X_4
+ 42 X_4^2 \,,
\\[2pt]
\mathcal{C} &\equiv
  480 X_1^2
+ 80 X_1 (2 X_2 + 3 X_3 + X_4)
+ 52 X_2^2
+ 64 X_2 X_3
+ 12 X_2 X_4
+ 36 X_3^2
+ 32 X_3 X_4
+ 23 X_4^2 \,.\notag 
\end{aligned}
\end{equation}
In the non-relativistic limit the thermal average of the annihilation cross section times the Møller velocity for the process \( Z'Z' \to gg \) is given by:
\begin{equation}
\begin{aligned}
& \left\langle  \sigma_{Z^\prime Z^\prime\rightarrow gg} v_{\mathrm{Møl}} \right\rangle  = \sigma_{Z^\prime Z^\prime\rightarrow gg}(\langle \sqrt{s} \rangle) ~\langle v_{\mathrm{Møl}} \rangle  \\
&  = \frac{\sqrt{2} m_{Z'}^6}{9\pi \Lambda^8} \Big( 96 X_1^2 + 12 X_2^2 + 8 X_3^2 + 8 X_3 X_4 + 3 X_4^2  + 16 X_1 (2 X_2 + 3 X_3 + X_4) 
+ 4 X_2 (4 X_3 + 3 X_4) \Big)\,.
\end{aligned}
\end{equation}

\subsection{\texorpdfstring{\(  gg\to Z^\prime Z^\prime \)}{} process}\label{freezeincalc3}

Due to crossing symmetry, the color-stripped amplitude squared
$\sum_{\rm pol}|\mathcal{M}|^2$, for the process
$g(k_1) + g(k_2) \rightarrow Z^\prime(p_1) + Z^\prime(p_2)$
is identical to Eq.~(\ref{eq.amplitudex}).
Therefore, in the high energy limit $s\gg 4m_{Z'}^2$, the expression of $\sum_{\rm pol}|\mathcal{M}|^2$ integrating over $t$ and summing over all color states is approximately 
\begin{equation}\label{eq:pairproductM}
    \sum_{\rm pol}\sum_{\rm col.}\int_{t_{\min}}^{t_{\max}} \left|\mathcal{M}\right|^2 \mathrm{d}t =  {2 s^5 \over \Lambda^8} F_1\left(\{X_j\}\right) +{\mathcal O}({m_{Z'}^2\over s}) \,,
\end{equation}
where the dimensionless function $F_1$, depending on the parameters $X_j$, is given by
\begin{equation}
\begin{aligned}
    F_1\left(\{X_j\}\right) &= \frac{2}{15} \Big[ 480 X_1^2 + 52 X_2^2 + 64 X_2 X_3 + 36 X_3^2 + 12 X_2 X_4 + 32 X_3 X_4 + 23 X_4^2 \nonumber \\
    &\quad + 80 X_1 (2 X_2 + 3 X_3 + X_4) \Big] \,.
\end{aligned}
\end{equation}
That is, as the high-energy limit the $Z'$-pair production cross section is 
\begin{equation}
\begin{aligned}
\sigma_{g g \to Z' Z'}(s)
&= \frac{1}{2!}\,\frac{1}{16\times16}\,\frac{1}{64\pi s}\,
    \frac{1}{P_{\rm cm}(s, 0, 0)^2}
    \sum_{\rm pol.} \sum_{\rm col.}
    \int_{t_{\min}}^{t_{\max}} \left|\mathcal{M}\right|^2 \, {\rm d} t
\\[3pt]
&\simeq  \frac{ s^3}{4096\pi \Lambda^8}
    F_1\!\left(\{X_j\}\right)\,.
\end{aligned}
\end{equation}

\subsection{$Z^{\prime}$--nucleon Scattering Cross-section }

We consider the $Z^\prime$–nucleon elastic scattering process
\begin{equation}
    Z'(p_1) + n(k_1) \rightarrow Z'(p_2) + n(k_2)\,,
\end{equation}
At the leading order, the matrix element of the gluon field strength tensor between nucleon states can be parameterized as~\cite{Bishara:2017pfq}
\begin{equation}
\langle n| \mathrm{Tr}\left[ G_{\mu\nu} G^{\mu\nu} \right] |n\rangle = -\frac{8\pi}{9\alpha_s} m_G \, \bar{u}_n u_n \,,
\end{equation}
where \( m_G \simeq 848~\mathrm{MeV} \) is the gluonic contribution to the nucleon mass in the isospin limit.
Consequently, the \( Z' \)-nucleon amplitude induced by the operator \( \mathcal{O}_1 \) is given by
\begin{equation}
\begin{aligned}
\mathcal{M}&=\frac{X_1}{\Lambda^4} \langle Z' | Z'_{\alpha\beta} Z'^{\alpha\beta} | Z' \rangle 
\langle n| \mathrm{Tr}[ G_{\mu\nu} G^{\mu\nu} ] |n\rangle 
\\& =-\frac{X_1}{\Lambda^4} \frac{8\pi}{9\alpha_s} m_G \, \bar{\psi}_n \psi_n \Big[  \, 4 (p_1 \cdot p_2) \left( \varepsilon(p_1) \cdot \varepsilon^*(p_2) \right) 
 - 4 (p_1 \cdot \varepsilon^*(p_2)) (p_2 \cdot \varepsilon(p_1)) \Big]\,.
\end{aligned}
\end{equation}
Using the Mandelstam variables
\[
s = (p_1 + k_1)^2 = (p_2 + k_2)^2 = (m_n + m_{Z'})^2 + m_n m_{Z'} v^2\,, \quad t = (p_1 - p_2)^2 = (k_1 - k_2)^2\,,
\]
with \( v \sim 10^{-3} \) being the typical galactic DM velocity, the spin- and polarization-averaged squared amplitude is given by
\begin{equation}
\frac{1}{2} \cdot \frac{1}{3} \sum_{\mathrm{spin}} \sum_{\mathrm{pol}} |\mathcal{M}|^2 =\frac{X_1^2}{\Lambda^8} \frac{8}{3} \left(4 m_n^2 - t\right)\left(6 m_{Z'}^4 - 4 m_{Z'}^2 t + t^2\right)\,.
\end{equation}
The total cross section is then given by
\begin{equation}
\sigma = \int_{t_{\min}}^{t_{\max}} \frac{1}{2} \cdot \frac{1}{3} \cdot \frac{1}{64\pi s} \cdot \frac{1}{P_{\mathrm{cm}}(s, m_{Z'}, m_n)^2} \sum_{\rm spin} \sum_{\rm pol} |\mathcal{M}|^2 \, dt = \frac{X_1^2}{\Lambda^8}\frac{256 \pi m_G^2 m_n^2 m_{Z'}^4}{81 \alpha_s^2 (m_n + m_{Z'})^2}\,,
\end{equation}
where $t$ ranges from
\begin{equation}
t \in [-\frac{m_n^2 m_{Z'}^2 v^2(4 + v^2)}{m_n^2 + m_{Z'}^2 + 2 m_n m_{Z'}(1 + v^2/2)},~0]\,.
\end{equation}

\section{Thermally Averaged Annihilation Cross-Sections }\label{app: thermally averaged annihilation cross sections}
A BSM species, labeled $3$, interacts with the SM through the $2\!\leftrightarrow\!2$ process
\begin{equation}
  1(p_1)+2(p_2)\;\longleftrightarrow\;3(p_3)+4(p_4)\,,
\end{equation}
where $p_i$ denote the four–momentum, particles 1 and 2 are SM bath states, and particle 4 may be either an SM or a BSM state. The time evolution of the number density $n_3$ is governed by the Boltzmann equation~\cite{Hall:2009bx},
\begin{equation}\label{eq.boltzmanna}
    \dot{n}_3 + 3 H(T) n_3 = R(T)\,,
\end{equation}
where $H(T)$ denotes the Hubble expansion rate, and $R(T)$ represents the interaction  efficiency, expressed as
\begin{equation}\label{eq:interactionterm}
\begin{aligned}
    R(T) &= \sum \frac{1}{\mathcal{S}_i}\frac{1}{\mathcal{S}_f} \int \mathrm{d}\Pi_1 \, \mathrm{d}\Pi_2 \, \mathrm{d}\Pi_3 \, \mathrm{d}\Pi_4 \, (2\pi)^4 \delta^{(4)}(p_1 + p_2 - p_3 - p_4) \\
    &\quad \times  |\mathcal{M}|^2 \,\left[ f_1 f_2 (1 \pm f_3)(1 \pm f_4) 
    -  \, f_3 f_4 (1 \pm f_1)(1 \pm f_2) \right]\,,
\end{aligned}
\end{equation}
with 
$$\mathrm{d}\Pi_i \equiv \frac{\mathrm{d}^4 p_j}{(2 \pi)^3} \delta_+(p_j^2 - m_j^2)$$
representing  the Lorentz-invariant phase-space measure for particle $i$. Here the $\pm$ signs account for quantum statistics, where $+$ ($-$) applies to bosons (fermions). The summation extends over all relevant production channels, incorporating contributions from all possible species of initial and final state particles, as well as their degrees of freedom from spin and color configurations. The factor ${\mathcal{S}_{i/f}}$ accounts for the symmetry factor associated with identical particles in the initial/final states, ensuring the correct counting of phase-space configurations. In the absence of Bose-Einstein condensation or Fermi-Dirac degeneracy, quantum statistical effects such as Pauli blocking for fermions and stimulated emission for bosons can be neglected. The distribution functions \( f_{\rm j} \) for particles \( i \) in kinetic equilibrium at temperature \( T \) can then be written as \( f_{\rm j} \approx e^{-(E - \mu_j)/T} \), where \( 1 \pm f_{\rm j} \approx 1 \) holds in this non-degenerate regime. Meanwhile, the distribution functions \( f_{\rm j~ eq} \) for particles in thermal equilibrium follow Maxwell-Boltzmann statistics, \( f_{\rm  j ~ eq} \approx e^{-E/T} \).

The interaction  efficiency $R(T)$ can be written in terms of number densities and the thermally averaged cross section as \cite{Young:2016ala}
\begin{equation}\label{eq:interactionrate}
    R(T)
    = 
      \frac{1}{\mathcal{S}_i}\,n_1^{(0)} n_2^{(0)}
      \left(
        \frac{n_1 n_2}{n_1^{(0)} n_2^{(0)}}
        - \frac{n_3 n_4}{n_3^{(0)} n_4^{(0)}}
      \right)
      \big\langle v_{\rm M\o l} \sigma \big\rangle \,,
\end{equation}
where $n_j$ denotes the particle number density for species $j$ in kinetic equilibrium,
\begin{equation}
    n_j
    = g_j \int \frac{{\rm d}^3 p}{(2\pi)^3}\,
      e^{-(E - \mu_j)/T}
    = g_j \frac{4\pi}{(2\pi)^3} \int_0^\infty
      e^{-(E - \mu_j)/T}\, p^2\, {\rm d}p\,,
\end{equation}
and $n_j^{(0)} \equiv n_j\big|_{\mu_j = 0}$ is the thermal–equilibrium value at temperature $T$, and $g_j$ is the internal degree-of-freedom, for the particle $j$.
The thermally averaged cross section is defined by
\begin{equation}
    \big\langle v_{\rm M\o l} \sigma \big\rangle
    \equiv
    \frac{g_1 g_2}{n_1^{(0)} n_2^{(0)}} 
    \int \frac{{\rm d}^3 p_1}{(2\pi)^3}
         \frac{{\rm d}^3 p_2}{(2\pi)^3}\,
         e^{-E_1/T}\, e^{-E_2/T}\,
         \left| \frac{\vec{p}_1}{E_1} - \frac{\vec{p}_2}{E_2} \right|
         \sigma(1 + 2 \to 3 + 4)\,.
\end{equation}
where $\sigma(1 + 2 \to 3 + 4)$ is the spin- and color-averaged cross-section for the process $1 + 2 \to 3 + 4$.  
The phase space integral methodology for $\langle v_{\mathrm{Møl}} \sigma \rangle$ can be found in \cite{Edsjo:1997bg}. In fact, the volume element can be written as 
\begin{equation}
    d^3 p_1\, d^3 p_2 = 4 \pi |p_1| E_1 \, d E_1 \, 4 \pi |p_2| E_2 \, d E_2 \, \frac{1}{2} d \cos \theta\,,
\end{equation}
which is then further simplified by redefining the integration variables as follows:
\begin{equation}
\left\{
\begin{array}{l}
E_{+} = E_1 + E_2\,, \\
E_{-} = E_1 - E_2\,, \\
s = m_1^2 + m_2^2 + 2 E_1 E_2 - 2 |p_1| |p_2| \cos \theta \,.
\end{array}
\right.
\end{equation}
As a result, the volume element becomes:
\begin{equation}
\frac{d^3 p_1}{(2 \pi)^3 2 E_1} \frac{d^3 p_2}{(2 \pi)^3 2 E_2} = \frac{1}{(2 \pi)^4} \frac{d E_{+} \, d E_{-} \, d s}{8}\,,
\end{equation}
and the integration region \( \{E_1 \geq m_1, E_2 \geq m_2, |\cos \theta| \leq 1\} \) transforms into:
\begin{equation}
\begin{aligned}
s &\geq (m_1 + m_2)^2\,, \\
E_{+} &\geq \sqrt{s}\,, \\
\left| E_{-} - E_{+} \frac{m_2^2 - m_1^2}{s} \right| &\leq  \frac{\left[s - (m_1 + m_2)^2\right]^{1 / 2} \left[s - (m_1 - m_2)^2\right]^{1 / 2}}{ \sqrt{s}} \sqrt{\frac{E_{+}^2 - s}{s}}\,.
\end{aligned}
\end{equation}
If one instead focuses on the evolution of $n_1$, Eq.~\eqref{eq.boltzmanna} should be rewritten as   
\begin{equation}\label{eq.boltzmanna1}
    \dot{n}_1 + 3 H(T) n_1 =  - R(T)\,,
\end{equation}
with the same definition of $R(T)$ as above. Moreover, if here particles 1 and 2 are identical, an additional factor of 2 is needed,   counting the annihilation of two identical particles per process. As is well known, this $ 1/2 $ factor  should cancel with the initial-state symmetry factor $\mathcal{S}_i$.

\section{Analytical Approximation of Freeze-in Calculation}\label{app:freezein}

In the freeze-in scenario, the feebly interacting massive particle (FIMP) resides in a cold hidden sector characterized by a negligible initial abundance,
\begin{equation}
n_{\rm FIMP}^{\rm init} \equiv \frac{g_{\rm FIMP}}{(2\pi)^3} \int \mathrm{d}^3 p \, f_{\rm FIMP}^{\rm init}(p) \approx 0\,,
\end{equation}
where $g_{\rm FIMP}$ denotes the number of internal degrees of freedom. Accordingly, the initial phase-space distribution satisfies $f_{\rm FIMP}^{\rm init}(p) \simeq 0$.  Through suppressed portal interactions induced by the operators \(\mathcal{O}_{1-4}\), \(Z'\) particles are gradually generated via out-of-equilibrium gluon fusion processes:
\[
g\left(p_1\right) + g\left(p_2\right) \rightarrow Z^{\prime}\left(k_1\right) + Z^{\prime}\left(k_2\right)\,,
\]
eventually yielding the observed relic abundance. The number density evolution depends on the Boltzmann equation, as given in Eq.~\ref{eq.boltzmanna}. The freeze-in scenario assumes $f_{\rm FIMP}(p) \ll 1$ at all times, which allows us to neglect the back-reaction terms. 

\begin{figure}[ht]
    \centering
    \begin{minipage}{0.495\textwidth}
        \centering
        \includegraphics[width=\textwidth]{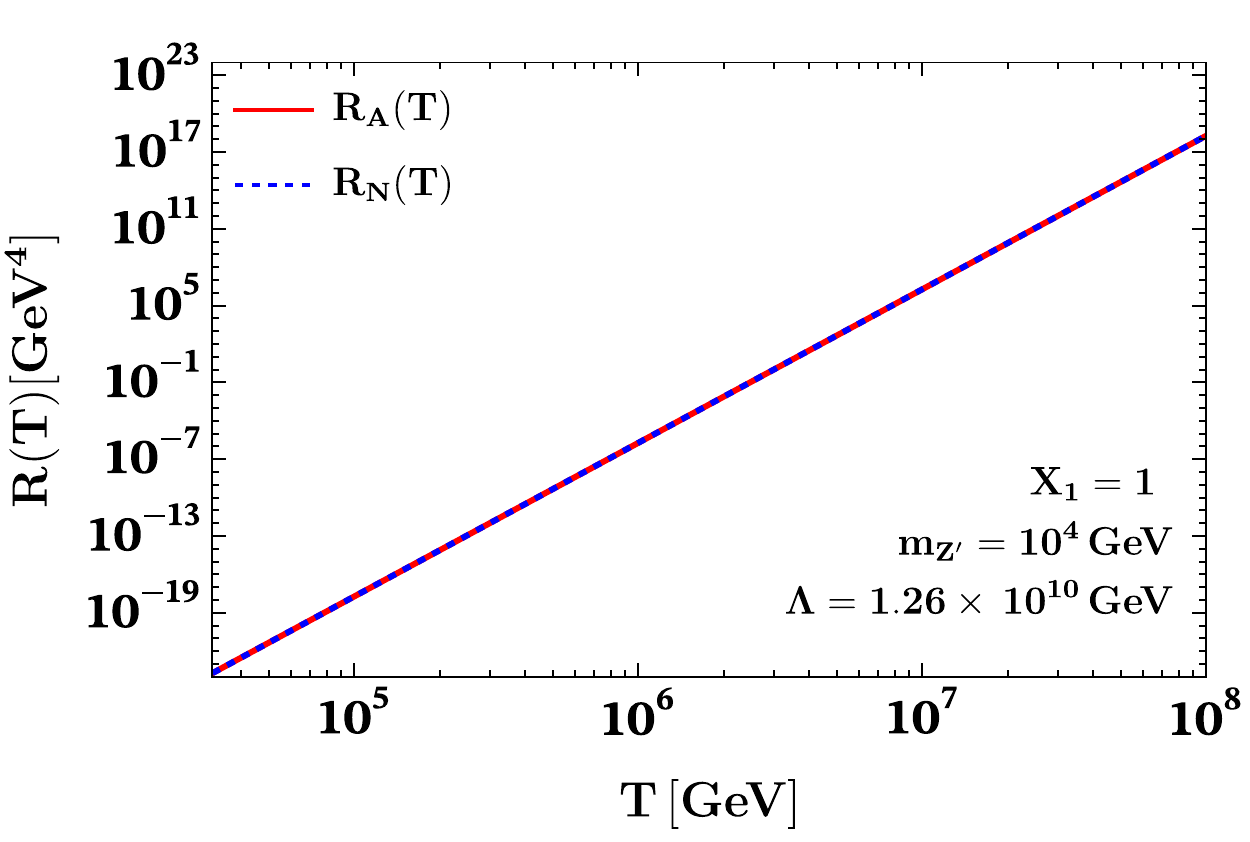}

    \end{minipage}\hfill
    \begin{minipage}{0.495\textwidth}
        \centering
        \includegraphics[width=\textwidth]{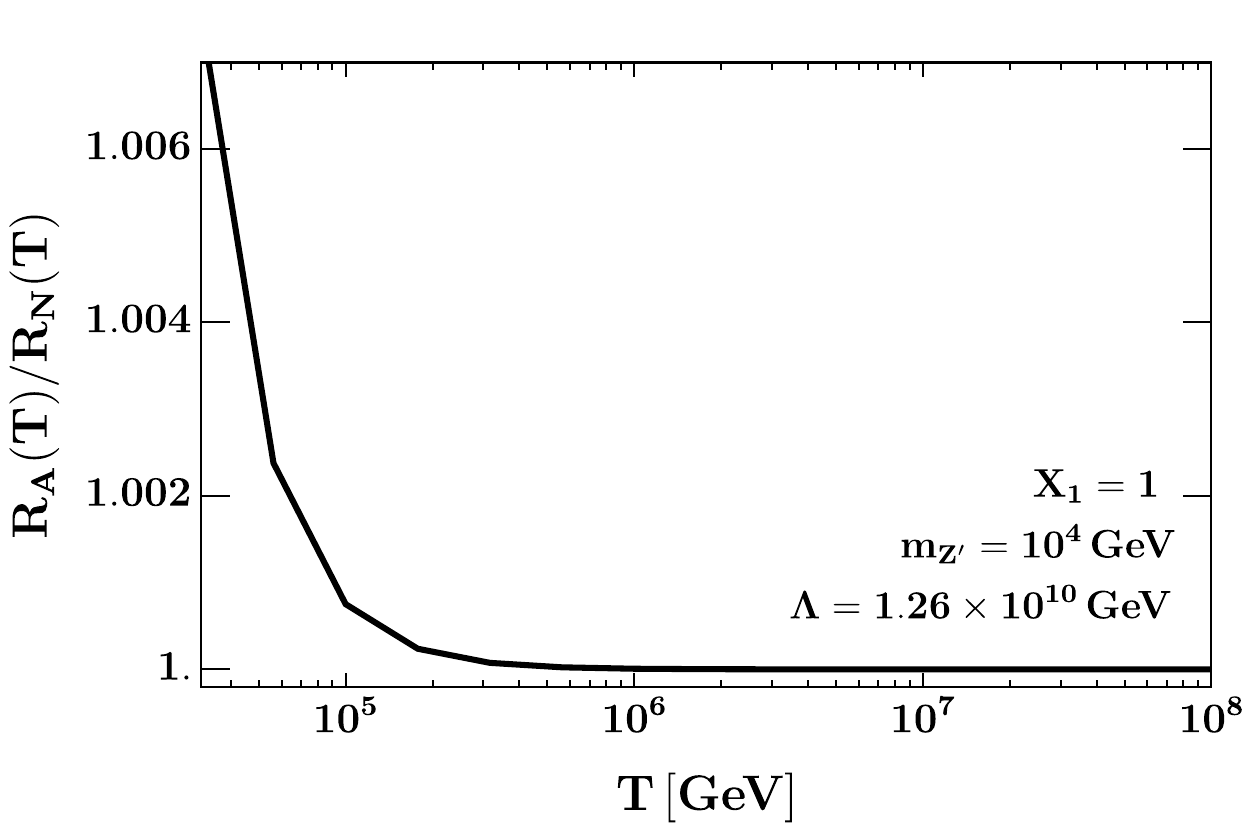}

    \end{minipage}
      \caption{{\bf Left} panel shows the interaction efficiency $R(T)$ computed   analytically (red solid line) and numerically (blue dashed line). {\bf Right} panel displays their ratio as a function of temperature $T$. These results are obtained for the benchmark scenario with $X_1 = 1$ ($F_1(\{X_j\}) = 64$), $\Lambda = 1.26\times10^{10}~\text{GeV}$, and $m_{Z'} = 10^4~\text{GeV}$, used in the main text.}
    \label{fig:RT}
 
\end{figure}

Therefore, the interaction  efficiency \( R(T) \) can be reformulated as an integral over the square of the center-of-mass energy \( s \) and the momentum-transfer \( t \), yielding~\cite{Hall:2009bx, Edsjo:1997bg}:
\begin{equation}\label{eq.RT}
R(T) \simeq 2\, \sum_{\rm pol}\sum_{\rm col.}  \frac{1}{\mathcal{S}_i}\frac{1}{\mathcal{S}_f} \frac{T}{512 \pi^5} 
\int_{s_{\rm min}}^\infty \mathrm{d}s \int_{t_{\rm min}}^{t_{\rm max}} \mathrm{d}t \, 
|\mathcal{M}_{gg \to z^\prime z^\prime}|^2 \frac{1}{\sqrt{s}} K_1\!\left(\frac{\sqrt{s}}{T}\right)\,,
\end{equation}
where \(K_1(x)\) denotes the modified Bessel function of the second kind, and  factor of 2 accounts for the two $Z^\prime$ produced in this process.
The integration limits for \(s\) and \(t\) are specified in Appendix~\ref{app:crosssection}. 
In the high-temperature limit (\(T \simeq \sqrt{s} \gg m_{i}\)), dimensional analysis yields a simplified expression for the \( t \)-integral:
\begin{equation}\label{eq.dimensionanaly}
2\,\sum_{\rm pol}\sum_{\rm col.} \frac{1}{\mathcal{S}_i}\frac{1}{\mathcal{S}_f} 
\int_{t_{\text{min}}}^{t_{\text{max}}} \left|\mathcal{M}_{gg \to z^\prime z^\prime}\right|^2 \, \mathrm{d}t 
\simeq  
\frac{s^{n' - 3} F_1(\{X_j\}) }{\Lambda^8}   + \mathcal{O}({ m_{Z'}^2 \over s } )\,,
\end{equation}
where \(n'\) denotes the dimension of the effective operator, and the functions \(F_1(\{X_j\})\) encode the coupling structure, depending on all  the coefficients \(X_j\). 
The explicit form of \(F_1(\{X_j\})\) is provided in Appendix~\ref{freezeincalc3}. 
Combining Eqs.~\eqref{eq.RT} and~\eqref{eq.dimensionanaly}, and noting that the \( s \)-integral admits a closed-form solution for \( n \in \mathbb{N} \),
\begin{equation}
\int_0^\infty \mathrm{d}s\, s^{(2n+1)/2} K_1\left( \frac{\sqrt{s}}{T} \right) 
= 4^{n+1} T^{2n+3} n!(n+1)!\,,
\end{equation}
one can simplify the interaction  efficiency \( R(T) \) for \( n' = 8 \) to
\begin{equation}
R(T) \simeq   \frac{5760 \, F_1(\{X_j\}) \, T^{12}}{\Lambda^8\,\pi^5}\,.
\end{equation}
The comparison is shown in the left panel of Fig.~\ref{fig:RT}, where the interaction  efficiency $R(T)$ is calculated using both the numerical integration method (blue dashed line) and the analytical approximation (red solid line). Here we assume that only the Wilson coefficient $X_1 = 1$ contributes (i.e., $F_1(X_j) = 64$), with $\Lambda = 1.26 \times 10^{10}~\mathrm{GeV}$ and $m_{Z'} = 10^4~\mathrm{GeV}$. The ratio of the analytical to numerical results is displayed in the right panel, demonstrating excellent agreement in the high-temperature regime, with the discrepancy rising at $T \to m_{Z'}$, as expected. This discrepancy has a negligible impact on the evolution of the number density for the UV-dominated freeze-in.

At last, using $\dot{T} \simeq - H T$, the Boltzmann equation~Eq.~\eqref{eq.Boltzmann_equation} can be rewritten in terms of the dimensionless yield $Y_3 \equiv n_3/s$ as
\begin{equation}
\frac{dY_3}{dT} \simeq -\frac{R(T)}{s H T}\,,
\end{equation}
where the entropy density $s$ and the Hubble parameter $H$ are given by
\[
s = \frac{2\pi^2}{45} g_*^S(T) T^3\,, \quad H = \frac{1.66\, \sqrt{g_*^\rho(T)}\, T^2}{M_{\mathrm{Pl}}}\,,
\]
with $M_{\mathrm{Pl}}$ denoting the Planck mass, and $g_*^{S,\rho}(T)$ representing the effective relativistic degrees of freedom for entropy and energy density, respectively. 
After freeze-in terminates, $Y_3$ becomes a constant, and its present-day value $Y_{3,0}$ is given by
\begin{equation}\label{eq.abundance}
\begin{aligned}
Y_{3,0} = Y_3^{\infty} &\simeq \int_0^{T_{\mathrm{RH}}} \frac{5760\, F_1(\{X_j\})\, T^8}{\Lambda^8\, \pi^5\, s H T} \, \mathrm{d}T \\
&\simeq \frac{11153.2\, F_1(\{X_j\})\, M_{\mathrm{Pl}}\, T_{\mathrm{RH}}^7}{\pi^7\, g_*^S(T_{\mathrm{RH}}) \sqrt{g_*^\rho(T_{\mathrm{RH}})}\,\Lambda^8\, }\,,
\end{aligned}
\end{equation}
where $g_*^\rho(T)$ and $g_*^S(T)$ are taken as being temperature-independent during the analytical approximation. In practice, we use their SM values at $T= T_\mathrm{RH}$ for freeze-in calculation.

\bibliographystyle{JHEP}
\bibliography{ref}
\end{document}